\documentclass[CRPHYS,Unicode,manuscript]{cedram}

\usepackage{soul}

\usepackage{subcaption}
\usepackage{bm}
\newcommand\unit[1]{~ \mathrm{#1}}

\title{Solid-liquid phase change in planetary cores} 

\author{\firstname{Ludovic} \lastname{Huguet}\CDRorcid{0000-0002-5532-6302}\IsCorresp}
\address{ISTerre, Université Grenoble Alpes, Université Savoie Mont Blanc, CNRS, IRD, Université Gustave Eiffel, 38000 Grenoble, France}
\address{Now at School of Earth and Environment, University of Leeds, Leeds LS2 9JT, UK}
\email[L. Huguet]{l.g.huguet@leeds.ac.uk}
\thanks{LH is supported by the European Research Council (ERC) under the European Unions Horizon 2020 research and innovation programme (grant number 716429). ISTerre is part of Labex OSUG@2020 (ANR10 LABX56).}

\author{\firstname{Quentin} \lastname{Kriaa}\CDRorcid{0000-0002-3518-1660}}
\address{Physics of Fluids Group, Max Planck Center for Complex Fluid Dynamics, and J. M. Burgers Centre for Fluid Dynamics, University of Twente, P.O. Box 217, 7500AE Enschede, The Netherlands}
\email[Q. Kriaa]{quentinkriaa@gmail.com}

\author{\firstname{Thierry} \lastname{Alboussière}\CDRorcid{0000-0002-3692-899X}}
\address{Université de Lyon, ENSL, UCBL, UJM, CNRS, Laboratoire LGL-TPE, France}
\email[T. Alboussière]{thierry.alboussiere@ens-lyon.fr}

\author{\firstname{Michael} \lastname{Le Bars}\CDRorcid{0000-0002-4884-6190}}
\address{CNRS, Aix Marseille Univ, Centrale Marseille, IRPHE, Marseille, France}
\email[M. Le Bars]{michael.le-bars@univ-amu.fr}
\thanks{MLB and TA are supported by the Programme National de Planétologie (PNP) of CNRS-INSU co-funded by CNES}

\ESM{Supplementary material for this article is supplied as a separate
archive available from the journal's
website under
article's URL
or from the author.}

\altkeywords{Cristallisation, changement de phase, écoulements diphasiques, noyaux planétaires, champ magnétique}
\begin{abstract}
The ubiquitous phenomena of crystallization and melting occur in various geophysical contexts across many spatial and temporal scales. In particular, they take place in the iron core of terrestrial planets and moons, profoundly influencing their dynamics and magnetic field generation. Crystallization and melting entail intricate multiphase flows, buoyancy effects, and out-of-equilibrium thermodynamics, posing challenges for theoretical modeling and numerical simulations. Besides, due to the inaccessible nature of the planetary deep interior, our understanding relies on indirect data from seismology, mineral physics, geochemistry, and magnetism. Consequently, phase-change-driven flows in planetary cores constitute a compelling yet challenging area of research. This paper provides an overview of the role of laboratory fluid dynamics experiments in elucidating the solid-liquid phase change phenomena occurring thousands of kilometers beneath our feet and within other planetary depths, along with their dynamic repercussions. Drawing parallel with metallurgy, it navigates through all scales of phase change dynamics, from microscopic processes (nucleation and crystal growth) to macroscopic consequences (solid-liquid segregation and large-scale flows). The review delves into the two primary planetary solidification regimes, top-down and bottom-up, and elucidates the formation of mushy and/or slurry layers in the various relevant configurations. Additionally, it outlines remaining challenges, including insights from ongoing space missions poised to unveil the diverse planetary regimes.\end{abstract}

\begin{altabstract}
Les phénomènes de cristallisation et de fusion se produisent dans divers contextes géophysiques à diverses échelles spatiales et temporelles. Ils prennent notamment place dans le noyau en fer des planètes telluriques et des lunes, où ils jouent un rôle primordial dans leur dynamique et dans la génération de leur champ magnétique. La cristallisation et la fusion se caractérisent par des écoulements multiphasiques complexes, des effets de flottabilité et une thermodynamique hors équilibre, difficilement accessibles à la modélisation théorique et aux simulations numériques. De plus, les profondeurs planétaires demeurent inaccessibles aux observations directes, et notre compréhension repose sur des données indirectes provenant de la sismologie, de la physique minérale, de la géochimie et du magnétisme. Par conséquent, les écoulements induits par un changement de phase dans les noyaux planétaires constituent un domaine de recherche à la fois passionnant et stimulant. Cet article donne un aperçu du rôle qu'ont les expériences de laboratoire en dynamique des fluides pour élucider des phénomènes de changement de phase solide-liquide se produisant à des milliers de kilomètres sous nos pieds ou dans d'autres planètes, ainsi que de leurs conséquences dynamiques. En établissant des parallèles avec la métallurgie, cet article navigue à travers toutes les échelles de la dynamique du changement de phase, des processus microscopiques (nucléation et cristallisation) à leurs conséquences macroscopiques (ségrégation solide-liquide et écoulements de grande-échelle). Nous décrivons en détail les deux principaux régimes de solidification planétaire, descendant et ascendant, ainsi que la formation de couches de solides interconnectées (``mush'') ou dispersées (``slurry'') dans les diverses configurations pertinentes. Enfin, cette revue décrit les défis restants, notamment dans le cadre de missions spatiales en cours qui dévoileront sans aucun doute la diversité des régimes planétaires.
\end{altabstract}

\begin{document}

\maketitle
\newpage

\section{Introduction}

Crystallization and melting are ubiquitous in various geophysical settings across large ranges of spatial and temporal scales. In most instances, solidification and melting are associated with subsequent buoyancy effects with profound dynamical consequences, leading to stable stratification or intense convective motions. Indeed, beyond latent heat and heat transfers, solidification and melting in geophysical settings are often intricately linked with species segregation due to the heterogeneity of the considered fluids. For instance, on our planet, the crystallization of ice at the poles increases the concentration of salt into the ocean water, hence playing a pivotal role in climate by driving part of the global ocean circulation. Diving towards the planetary deep, the significance of the local solid-liquid phase change of the mantle is revealed through volcanic activity. Partial melting within the lithosphere also shapes the composition of the crust and contributes to the formation of stable continents. In the early epochs of Earth's history, more substantial melting occurred within the mantle, driven by the energetic processes of accretion and the presence of short-lived radiogenic elements, resulting in a partial or global magma ocean setting the initial thermochemical state of today’s solid mantle. Still going deeper, the inner part of the Earth’s core has been solidifying over the last billion years or so, sustaining the Earth's magnetic field generation. Similar phase changes involving water, silicate, and iron are actually ubiquitous on all terrestrial planets and icy moons.

\subsection{Phase change and planetary magnetic fields}
 
The planet of reference for magnetic field production is the Earth, whose magnetic field is most likely produced by {dynamo} in its liquid core. This process describes how the motions of an electrically conducting fluid produce a magnetic field. The coupling between magnetism and fluid motions leads to the presence of electrical currents that feed the magnetic field but are constantly subject to ohmic dissipation. As a result, dynamo action can be sustained only as long as the electromotive action of the flow is sufficiently large compared to ohmic dissipation, which translates as a condition on the magnetic Reynolds number (e.g., \cite{cog1999})
\begin{equation}
    \label{eq:Rem_intro}
    Re_m = \frac{UL}{\eta_m} \gg 1 \ ,
\end{equation}
where $U$ and $L$ are respectively the characteristic velocity scale and length scale of the flow, and $\eta_m$ is the magnetic diffusivity. This necessary yet insufficient requirement places constraints on the vigor of the flow. Then, a burning question when observing a dynamo-generated magnetic field is: how do planetary cores sustain fluid motions that drive a dynamo?
\par
Three contributions are usually mentioned that participate in sustaining the Earth's magnetic field: \textit{thermal convection} due to the secular cooling of the planet, and possibly to internal heating through radioactive decay; \textit{crystallization} of the core at the center that leads to chemical segregation between pure iron (that solidifies on the solid inner core) and lighter elements that rise in the liquid outer core, feeding compositional convection; \textit{latent heat} release during crystallization that warms the fluid near the inner core, hence boosting the thermal convection. These three elements have proven important in driving the Earth's dynamo, and the last two of them are directly related to core crystallization. Earth's energy budget makes it very unlikely to sustain a dynamo without contribution from crystallization, hence questioning the model of a convective Earth dynamo before the start of the inner core growth (e.g., \cite{lfncs2022}). In addition to these three sources, mechanical driving processes (libration, precession, tidal deformation) can provide an alternative source of energy for generating magnetic fields in planetary cores \cite{km1998,lwkcl2011,dsn2011,lcl2015,lbblnst2022}.
\par
Some small Earth-like planets show evidence of past (Mars, Moon) or present (Ganymede, Mercury) global magnetic field (e.g., \cite{brs2015} or \cite{bsvvsr2022} for interior structure). These observations have questioned the scientific community, as long-term, global magnetic fields seem difficult to sustain in such small cores. In contrast to the Earth's bottom-up core crystallization, a regime of top-down crystallization might take place on small planets. Associated flows may sustain a dynamo, but their complexity has been up-to-now only remotely tackled.

Planetary cores are very particular geophysical objects, as no direct observation can be made. Our knowledge of the deepest part of planets relies on indirect observations from geochemistry (for composition), geodesy and gravity measurements (for internal structure, mass, moment of inertia, libration), seismology (for internal structure, density, velocity, and state), and magnetism. Since the early 20th century, the image of the Earth's core seen by the seismology shows stratified layers (E$^\prime$ layer and F-layer) \cite{icl2018,wipt2023} acting against the well-mixed state of the liquid core. Seismological studies have also revealed a complex structure inside the Earth's inner core, including east-west hemispheric dichotomy, innermost inner core, radial or lateral variation of elastic anisotropy, and attenuation \cite{d2014,wipt2023}. Planetary seismology on the Moon \cite{wlgwl2011,ggcl2011,garcia2019lunar} and on Mars \cite{skblgcddgho2021,sdrxhgliblo2023,ilddkrksabo2023} have also demonstrated the presence of a liquid core and some partially solid part at the top of the core or the bottom. The advance of satellite observations and planetary space missions has enabled us to better constrain the structure, intensity, and secular variation of planetary magnetic fields. The precision of geomagnetic measurements has even allowed us to infer the dynamics at the top of Earth's core, which provides key information on the state and the velocity of the fluid iron motions. However, in contrast with atmospheric and oceanic observations, the lack of direct measurements challenges our knowledge of the state and dynamics of planetary cores and makes it necessary to challenge the output of phase-change experiments and models with the available data.

\subsection{From the casting industry to planetary cores}
This review focuses on solid-liquid phase change within planetary cores and its dynamic consequences. In this context, the phase change is envisaged through its planetary, large-scale significance. But its intricate nature, encompassing both macroscopic and microscopic scales, is always present. At the atomic or molecular level, energy barriers must be overcome to induce a change in atomic arrangement. Diffusion processes operating at microscale influence the morphology of the liquid-solid interface, resulting in the formation of isolated crystals or dendritic structures, possibly modified by convection. The thickness of the region undergoing crystallization then represents the mesoscale, potentially involving compaction and other associated effects. While most studies focus on specific scales, it remains challenging to integrate all of them into a single numerical model. Consequently, experimental approaches hold significant value as they naturally capture all the complexity of phase change, albeit with the challenge of disentangling the individual contributions.

Many of our ideas on crystallization in planetary cores originate from the study of metallurgy (e.g., \cite{f1974,kfr2023}). Yet the conditions are very different: the expected growth rate of crystallization of the inner core -- $V \sim 10^{-10}$~m~s$^{-1}$ -- is unseen in the iron casting where the mm~s$^{-1}$ is more typical. Similarly, the temperature gradient in the core -- $G \sim 10^{-3}$K~m$^{-1}$ -- is much smaller than usual values in metallurgy. A consequence is that heuristic laws valid in metallurgy may not apply to the core. Fortunately, we can still rely on laws based on physical arguments. For instance, the primary dendrite spacing is found to scale proportionally to $V^{-1/4} G^{-1/2}$, based on diffusion length scales and the radius of curvature of the tips of the dendrites. This kind of scaling has been applied to obtain estimates of dendrite spacing and grain size in the inner core, reaching orders of magnitude of tens of meters \cite{dab2007}, well beyond day-to-day experience. 

Many factors in the planetary interiors do not have their counterparts in metallurgy. Compaction due to the deformation of crystallizing material under its weight or other stresses is a key factor in cores but is hardly studied in material processing. The effects of rotation and a background magnetic field are also potentially very important in planetary interiors. Magnetic fields have been investigated in the framework of steel casting, but not in the same view. They have been used to stir the fluid (traveling magnetic fields) or to brake fluid motion (DC magnetic fields), depending on the configuration and the desired final properties. At the scale of a solidifying zone, in the presence of a temperature gradient, thermoelectric currents can develop at the liquid-solid interface, acting as a short-circuited thermocouple. These currents have been measured using small electric potential probes \cite{lielausis84}. When a constant magnetic field is applied, the thermoelectric currents produce Lorentz forces and drive a flow that can compete and reverse the flow induced by compositional buoyancy \cite{alboussiere91,lehmann98}. All the ingredients for similar effects exist in metallic cores but have received little attention so far in this context.

\subsection{Outline}
This review is organized as follows. In section \ref{sec:cry_planet}, we first describe the state-of-the-art of our thermodynamical understanding of planetary cores' phase change and the variety of solidification regimes. This leads us to define two main scenarios, whose dynamics are described in depth in the following sections: the formation of a mushy layer detailed in section \ref{sec:mushylayer}, and the formation of a slurry layer detailed in section \ref{sec:slurryLayer}, respectively. Mushy layers correspond to situations where the solid forms a continuous, yet possibly porous matrix attached to a boundary. Slurry layers correspond to mixtures of fluid and detached grains in suspension, no matter what their concentration. The final section \ref{sec:conclusion} details the main remaining open questions and the upcoming challenges related to ongoing space missions. It also briefly describes phase change applications in other planetary settings, which could be each the subject of a long review.

\section{Crystallization in planetary cores}
\label{sec:cry_planet}
\subsection{\label{subsec:stateOfCore}Composition and state}

The solidification scenario of a planetary core depends on its liquidus temperature (also referred to as the solidification temperature or the melting temperature) and on its effective temperature profile (geotherm), which both depend on the pressure profile and core composition.

Planetary cores are primarily composed of iron\footnote{more specifically, of an iron-nickel alloy}, plus some light elements which could include sulfur, oxygen, silicon, hydrogen, and carbon \cite{hlh2013,brs2015}. Mineralogical studies of these various iron alloys under the pressure and temperature conditions of planetary cores allow for the determination of their density, melting temperature, thermal conductivity, and compressibility, depending on their composition and whether they are in a solid or liquid state. The melting temperature and thermal conductivity help determine when the first solid will form due to the thermal evolution of the core. The liquid compressibility and the slope of the liquidus determine whether crystallization will begin at the core's center or surface. The density difference between the liquid and the solid (positive or negative) determines the dynamics of liquid/solid segregation. All these parameters have been measured in various experimental studies for different alloys, such as Fe-Si \cite{kv2018}, Fe-S \cite{msfmrpb2007,bw2011}, Fe-S-O \cite{evswdh2019,pldf2018}, Fe-Si-H \cite{hthoouo2022}, Fe-Ni-S \cite{ll2020,pl2022}, Fe-Ni-Si \cite{ioh2021,kpskbsgkm2019}, and Fe-Ni-H \cite{stotnfh2014}.

\begin{figure}
\centering
\includegraphics[width=1\textwidth]{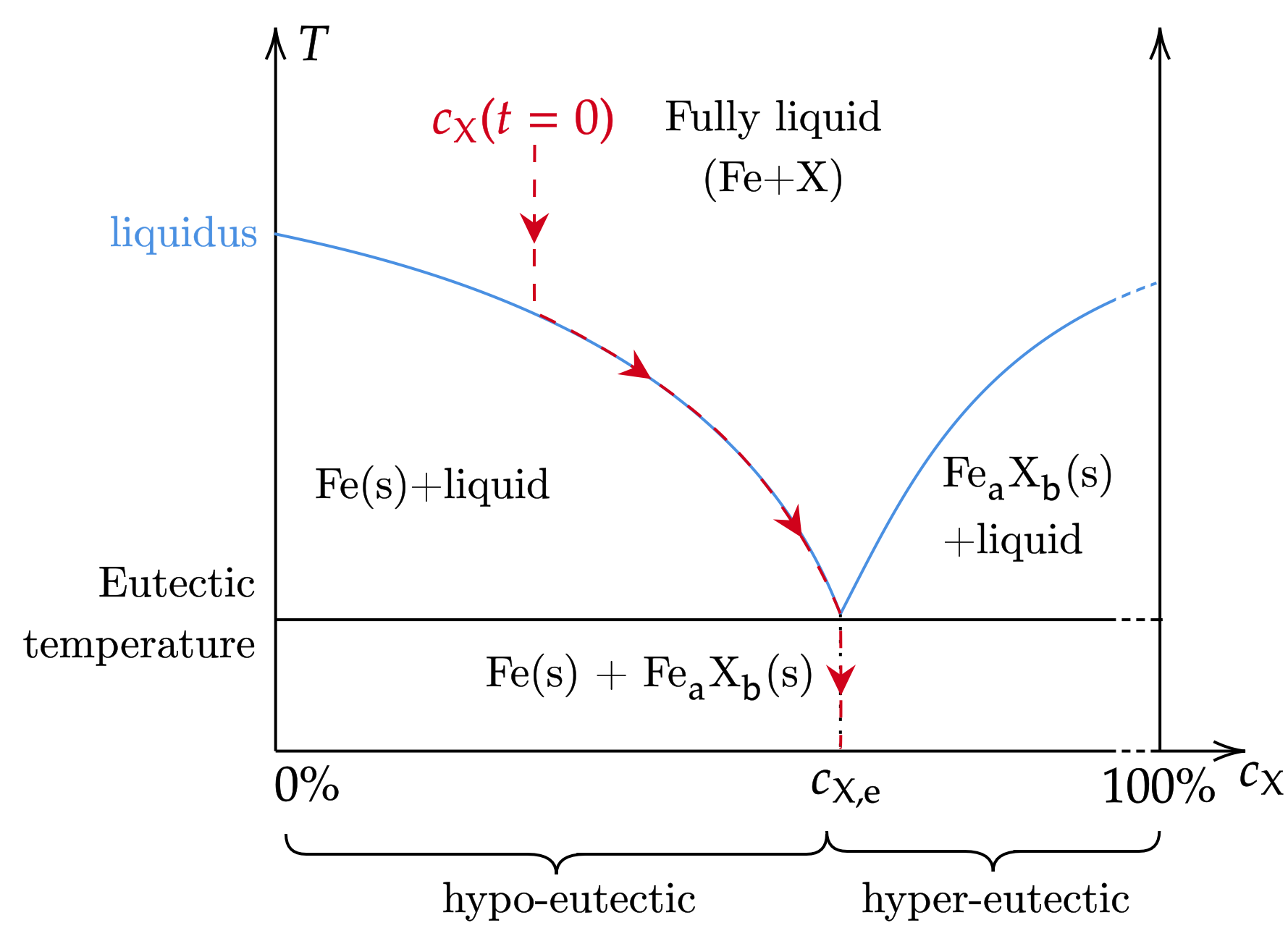}
\caption{Simplified sketch of the binary phase diagram of iron (Fe) and a light element denoted `X', showing the state of the core as a function of the X mass fraction $c_\mathrm{X}$ and of temperature. The solid blue line corresponds to the liquidus, and the solid dark line is the solidus. The composition of the eutectic is denoted $c_\mathrm{X,e}$. The integers $a$ and $b \neq 0$ in subscript parameterize the composition of the solid forming on the X-rich side of the eutectic. For simplicity, we focus in section \ref{subsec:scenariosSolidification} on a core lying on the iron-rich side of the eutectic (hypo-eutectic). In that case, the dashed red line shows a typical evolution of the state of the liquid core assuming thermodynamic equilibrium, for an initial mass fraction $c_\mathrm{X}(t=0)<c_\mathrm{X,e}$ and an initial temperature above the liquidus.}
\label{fig:binaryPhaseDiagramFeX}
\end{figure}

In the canonical framework, the core is assumed to be well-mixed during its accretion, leading to a uniform bulk composition before the crystallization starts \cite{n2015}. The liquid mixture is then described by a Fe-X alloy where the element X stands for a light element. In addition, a partition coefficient equal to 0 is usually considered, meaning that the iron solid phase is pure. In that case, the state of the core is described by the typical binary phase diagram sketched in Figure~\ref{fig:binaryPhaseDiagramFeX}. Such a diagram is given for a constant pressure, hence for a fixed depth in the core. Mineral physics studies provide the eutectic composition and temperature, and the Clayperon slope for a wide pressure range. Parameterized equations of state are built to determine which phase exists for all compositions, temperatures, and pressures. Over time, the core cools down, so the state of the core is generally expected to go from fully liquid to partially solid and, ultimately, fully solid. It is apparent in Figure~\ref{fig:binaryPhaseDiagramFeX} that the bulk X mass fraction in the fully liquid state controls the composition of the crystals that form. Cores on the iron-rich side of the eutectic (hypo-eutectic) $c_\mathrm{X}<c_\mathrm{X,e}$ form pure iron crystals; on the opposite, the X-rich cores (hyper-eutectic) $c_\mathrm{X}>c_\mathrm{X,e}$ form Fe$_\mathrm{a}$X$_\mathrm{b}$ crystals, which are generally considered positively buoyant.

\subsection{Dynamical scenarios of solidification}
\label{subsec:scenariosSolidification}

Figure~\ref{fig:binaryPhaseDiagramFeX} offers a simple, local picture of core crystallization solely based on thermodynamics. But in practice, core crystallization is a dynamic process due to the presence of buoyancy effects. Since the liquid Fe-X mixture contains a light element, pure iron crystals are denser than the liquid, and are therefore negatively buoyant, while Fe$_a$X$_b$ crystals are positively buoyant. The segregation of the newly solid phase and the remaining liquid phase determines the dynamical evolution of the cores, including the growth of a solid layer. Moreover, this segregation may lead to the formation of convective or stably stratified fluid zones. Fluid motions driven by buoyant parcels of fluid or by buoyant crystals are thought to generate enough energy to sustain a dynamo action \cite{b1963,lb1995,dp2018}.

The pathways of crystallization and segregation depend on two aspects. First, it depends on how the density of the crystals compares with that of the ambient. Second, crystal migration depends on exactly where solidification happens. For illustration purposes, we focus here on an iron-sulfur core on the iron-rich side of the eutectic, i.e. with a sulfur concentration $c_\mathrm{S}<c_\mathrm{S,e}$ and we describe two scenarios corresponding respectively to the evolution of the Earth and of a small Earth-like planet (or moons) like Ganymede. More crystallization scenarios can be found in \cite{brs2015}. In all scenarios, the ultimate state of the core is to be entirely solid (for small cores such as asteroids) or partially solid (for larger cores such as those of the Moon, Mercury, and Earth).

\begin{figure}
\centering
\includegraphics[width=\textwidth]{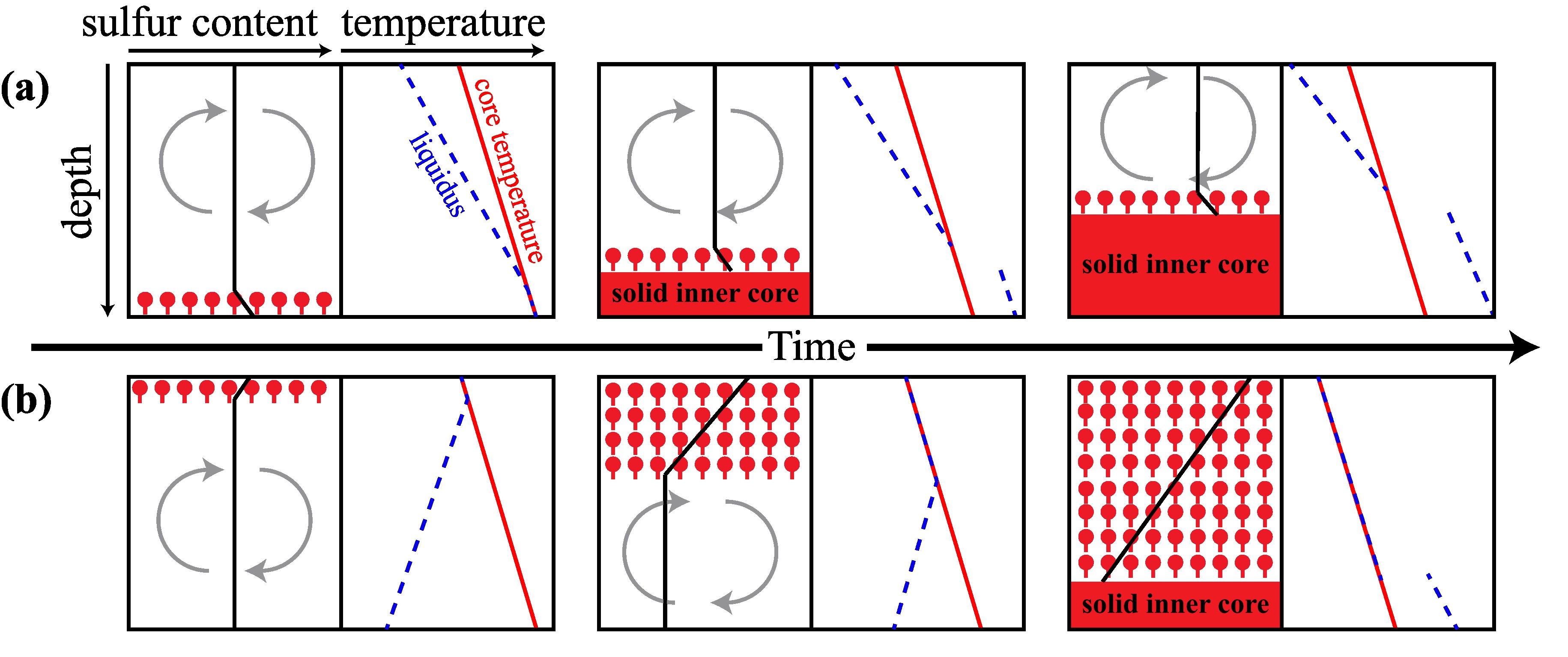}
\caption{Two dynamical scenarios of solidification of an iron-rich, sulfur hypo-eutectic core: (a) bottom-up solidification in an Earth-like planet; (b) top-down solidification in a smaller Earth-like planet. Figure adapted with permission from Breuer et al.~(2015) \cite{brs2015}, under the \href{http://creativecommons.org/licenses/by/4.0/)}{CC BY 4.0 license}. }
\label{fig:sampleBreuer2015}
\end{figure}

Figure~\ref{fig:sampleBreuer2015}a shows the typical bottom-up regime of solidification of the Earth \cite{j1953}. As the core cools down, its temperature eventually crosses the liquidus curve at its center because the slope of the core temperature is steeper than the liquidus. Consequently, solid crystals first appear at the planet center, as sketched by red circles. At the bottom of the crystallizing region, an iron solid layer grows by the formation of a solid matrix or by the accumulation of free iron crystals (see Section~\ref{subsec:mushyslurry}). In the meantime, the residual liquid depleted in iron is lighter than the bulk Fe-S mixture, so it rises in the core and nourishes compositional convection above the solid layer (gray arrows in Figure~\ref{fig:sampleBreuer2015}a).

Consider now the top-down freezing of small Earth-like planets like Ganymede \cite{had2006}, as sketched in Figure~\ref{fig:sampleBreuer2015}b. In such planets, studies \cite{fbf1997,cb2007,bw2011,rvvmd2011,dr2015} suggest that the slope of the liquidus as a function of depth is either steeper than the core temperature or even negative \cite{rvvmd2011,dr2015}. Therefore, as the core cools down its temperature profile first crosses the liquidus at the core periphery, where the first crystals appear \cite{brs2015,rbs2015,rbs2018} (see the red circles in Figure~\ref{fig:sampleBreuer2015}b). Iron crystals might grow from the core-mantle boundary or spontaneously nucleate freely in the bulk. The formation of the first crystals on the surface or in the bulk will depend on physical properties (wettability, roughness, nucleation rate \cite{hvhw2018}), and dynamical processes (growth rate, constitutional supercooling layer \cite{dab2007}, collisional breeding \cite{so1994}). We will deepen the conditions of crystallization in the section \ref{subsec:mushyslurry} and \ref{subsec:nucleation}. As they are denser than the ambient liquid, both cases end up with unstable crystals that settle due to gravity. The remaining light-element-rich liquid is lighter than the ambient, so it rises. The crystallization region extends from the core-mantle boundary to a lower limit where the crystals remelt because of the increasing ambient temperature. As a first-order estimate, this location is given by the core temperature matching the liquidus \cite{brs2015,rbs2015,rbs2018}. This region of solidification is called the {snow zone,} and the whole phenomenon of crystallization-settling-remelting is often referred to as {iron snow}. This snow zone is often thought to be at thermodynamic equilibrium, leading to the assumptions of fast crystallization and remelting compared to the crystal dynamics \cite{dp2018}. For the sake of simplicity, the departure from equilibrium in the snow zone is assumed to be negligible, even though supercooling can modify the snow dynamics in some experiments \cite{hld2023} (see Section \ref{subsec:bulkCrystallisationXP}). However, quantifying the departure from equilibrium requires releasing the previous hypotheses, making the theoretical model complex to handle \cite{l1992,wdj2023}. At the lower edge of the snow zone, the molten snowflakes form a liquid mixture that is richer in iron than the ambient liquid, all the more as it is concentrated in molten snowflakes, all the less as it dilutes with the ambient. This {molten snow} is therefore negatively buoyant, so it sinks towards the center of the core and nourishes compositional convection -- see \cite{c2015} for a numerical study. Eventually, the core temperature becomes lower than the liquidus everywhere, so that snowflakes accumulate at the core center and pile up as a growing inner core.
\par
Both scenarios described here are based on sketches and idealized hypotheses: they now need to be confronted with the reality of lab experiments.

\subsection{Mushy or slurry layers}
\label{subsec:mushyslurry}

\begin{figure}
\centering
\includegraphics[width=\textwidth]{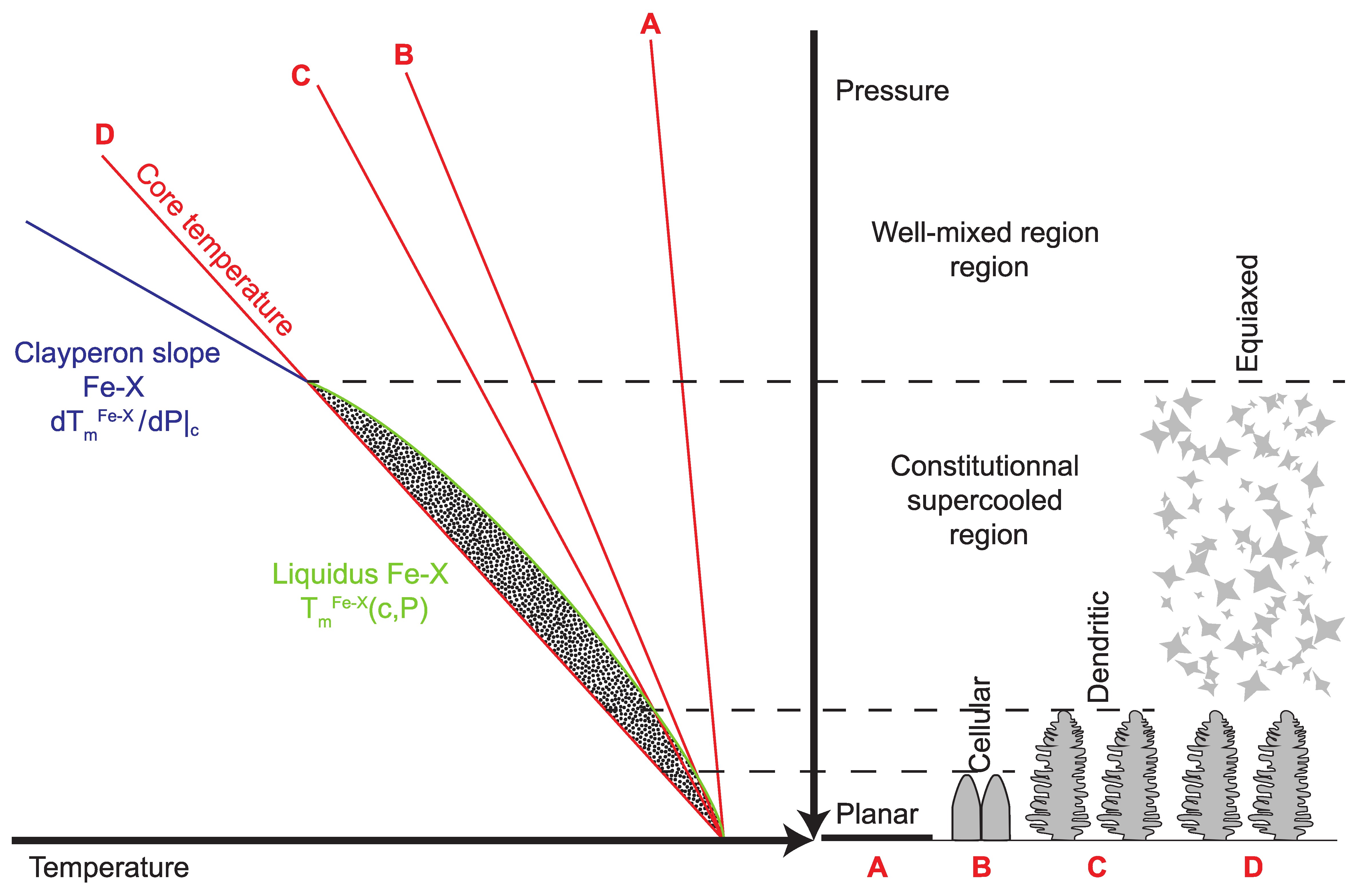}
\caption{Schematic view (for the Earth's core) of the core temperature profile (very close to the adiabatic profile), liquidus profile, and Clayperon slope, associated with a stable (A) or unstable (B-D) solidification front. An unstable front of solidification will encounter cellular (B), dendritic (C), or equiaxed growth (D). The thickness of the constitutional supercooled region (black dashed lines) depends on the core temperature and the liquidus profile, which both depend on the growth rate and composition.}
\label{fig:frontsolidification}
\end{figure}
During the crystallization of a binary mixture, the equilibrium temperature at the interface between the solid and liquid is equal to the liquidus temperature. However, the growth of the solid by crystallization requires a slight departure from this equilibrium, leading to the formation of supercooled regions ahead of the solidification front (i.e. the liquid temperature is below the freezing point). In addition, the composition of the liquid and the solid differ in terms of a light element composition depending on its solubility (generally very small for light elements (S, C, H, O), except for Si). Near the solidification front, the liquid is enriched in light elements, leading to the formation of a compositional gradient. Therefore, the fluid ahead of the front of solidification is often in a constitutionally supercooled state \cite{rc1953}. The morphological stability of the interface, which depends on both the temperature and composition gradients, has been extensively examined theoretically and experimentally \cite{ms1964,kfr2023}. In the conditions of planetary crystallization, the pressure gradient also plays a major role in the front stability, as the core and liquidus temperatures are pressure-dependent \cite{splm2005,dab2007} (Figure~\ref{fig:sampleBreuer2015}). If the interface is morphologically unstable, the front of solidification will encounter cellular growth, dendritic growth, or equiaxed crystal growth when considering an increasingly large zone of the constitutional supercooled layer \cite{kfr2023,c2002} (Figure~\ref{fig:frontsolidification}). In planetary cores, the stability of the front of solidification depends on the liquidus and Clapeyron slope (Figure~\ref{fig:frontsolidification}). The liquidus slope $\left. \frac{\partial T_m}{\partial c}\right|_P$ is negative or positive depending on whether the liquid composition is hypo-eutectic or hyper-eutectic (Figure~\ref{fig:binaryPhaseDiagramFeX}). The Clapeyron slope $\left. \frac{\partial T_m}{\partial P}\right|_c$ is mainly positive for iron alloys at high pressure, yet, for Fe-S at an intermediate pressure, the Clapeyron slope is negative \cite{clh2008,pldf2018,pl2022}. In addition, the solidification rate and advective transport of the solute in the liquid play a key role in controlling the structure of the thermal and chemical fields near the solidification front. While crystallization is very slow and is of the order of one millimeter a year \cite{lpl2001,n2015}, advective transport in the liquid can be several orders of magnitude larger \cite{habdll2016}. For the Earth, the theoretical analysis of the stability of the solidification front, at the conditions of the inner core boundary, suggests an unstable front of solidification \cite{flr1981,splm2005,dab2007}.

\begin{figure}
\centering
\includegraphics[width=\textwidth]{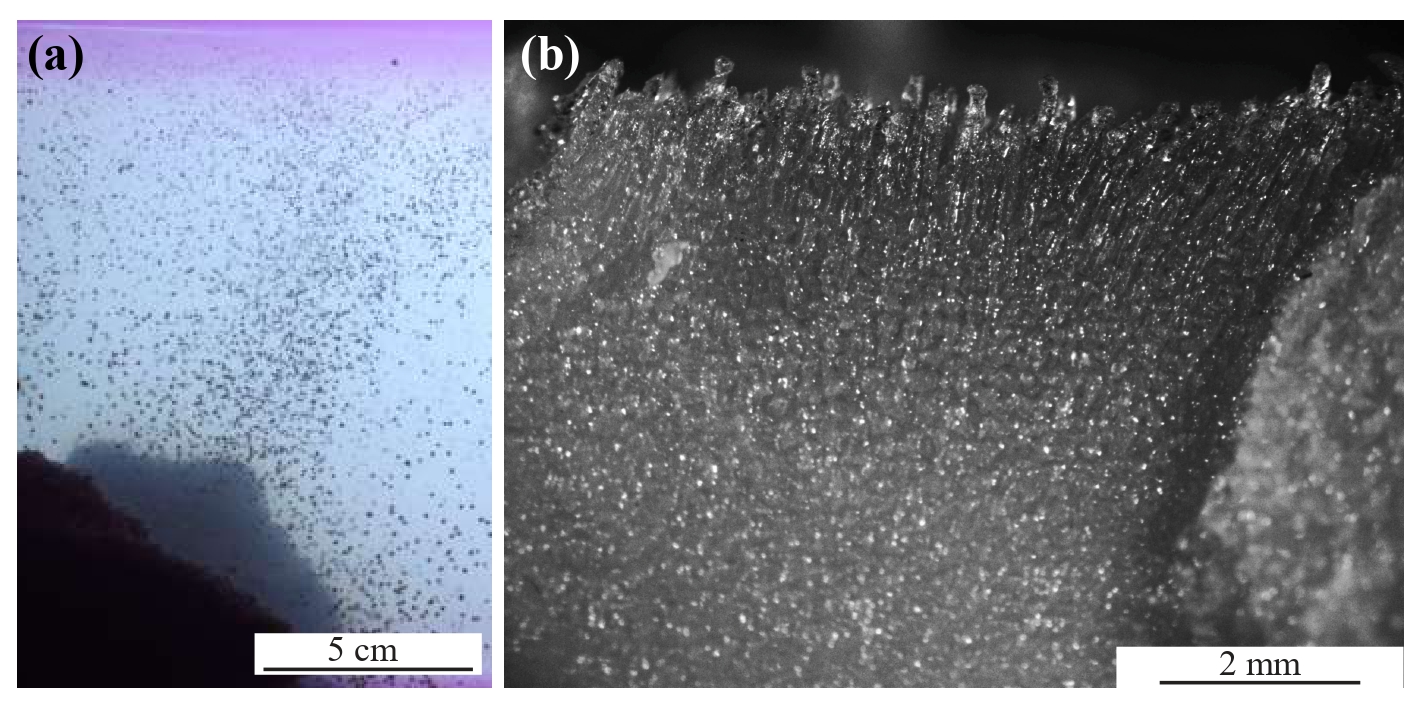}
\caption{(a) A slurry layer in a tank cooled from above. Equiaxed crystals form near the top boundary, settle, and accumulate at the bottom of the tank. (b) Mushy layer with dendrites growing from the bottom of a tank cooled from below \cite{h2014,habdll2016}. Both cases used ammonium chloride solutions as an analog to the Fe-X mixture \cite{jhus1966}.}
\label{fig:dendritevsequiaxed}
\end{figure}

Figure~\ref{fig:dendritevsequiaxed} shows two experiments of equiaxed and dendritic crystallization using analog material, like metallurgy studies. Equiaxed crystal growth leads to the formation of a two-phase region in which free crystals nucleate and grow, and the surrounding liquid can be convectively stable or unstable: this is a so-called \textit{slurry layer} (Figure~\ref{fig:dendritevsequiaxed}a). Alternatively, dendritic growth leads to the formation of a two-phase region in which dendrites coexist with an interstitial liquid that can also be stable or unstable: a so-called \textit{mushy layer} (Figure~\ref{fig:dendritevsequiaxed}b). In a mushy layer, primary dendrites grow mainly in the same direction. Both regimes can co-exist for the same conditions; mushy layers preferentially grow from a substrate, and slurry layers exist if enough nucleation sites are present.
Hence, by analogy with metallurgy, the crystallization conditions at the inner core boundary may lead to the formation of a mushy layer \cite{flr1981,splm2005,dab2007,habdll2016} or of a slurry layer \cite{splm2005,wdj2021}. The presence of a slurry layer depends on the thickness of the constitutional supercooled layer \cite{splm2005}, and the ability to produce nucleation sites in the bulk of this layer (see Section~\ref{subsec:nucleation}) \cite{rl1983,splm2005}. The stability of the front of solidification in other planetary cores remains poorly known even though the presence of dendrite crystallization has been suggested in asteroids based on observations in metallic meteorites \cite{hs1992,ch2006}, and the presence of an iron snow regime (slurry layer) has been suggested at the top of the core of small moons or planets \cite{had2006,rbs2015} (see Figure~\ref{fig:sampleBreuer2015}b).

\subsection{Nucleation paradox}
\label{subsec:nucleation}
Crystallization has been extensively studied in the literature (e.g., \cite{kfr2023}), given its importance in industrial processes (metallurgy, casting, ...). It is divided into two steps: nucleation followed by crystal growth. While the second step has also been widely studied in the context of planetary cores, the nucleation step has been largely ignored \cite{splm2005,n2015}. For liquid metals, classical nucleation theory \cite{kfr2023,c2002} requires either a very large supercooling in a homogeneous environment or a wettable surface reducing the supercooling. Experiments in various pure metals or alloys \cite{t1950,gsw1965,c2002,smm1987} have shown that the critical supercooling is quite large compared to the melting temperature, i.e. about 20\% of $T_m$. To explain the stability of the supercooled metals, Frank suggested that local structures exist in supercooled liquids \cite{f1952}. This hypothesis was later confirmed by identifying atomic structures with a degree of icosahedral short-range order in experiments in alloys \cite{klghrrrr2003} or in iron \cite{shsbh2002,gw2008,imkn2009}. These icosahedral short-range order structures may lead to the formation of stable or metastable quasi-crystals \cite{sbgc1984}, which would preferentially grow instead of crystalline phase \cite{klghrrrr2003}. The undercoolability (meaning the ability to remain stable for a supercooled liquid) of metal alloys then depends on the stability of these local structures in the supercooled liquid, which would ultimately depend on the presence of other phases than the iron \cite{hssbchh2004}. These short-range order structures have been suspected to play a major role in determining the melting temperature and the viscosity of high-pressure liquids \cite{br2015}. However, the experimental detection of such structure in a supercooled liquid at high pressure and temperature is extremely difficult. Still, the detection of footprints, such as the twining of dendrites in Al alloys \cite{kjr2013}, might be plausible. One may think that the advance of electromagnetic levitation containerless experiments \cite{kglhhkrrr2002} in various environments including microgravity \cite{gsbhvkgphk2021,mbggghkkvvo2023} along with laser shock experiments \cite{lwlxtzzxgs2020,khabbbbccgco2022}, which already provide physical properties of iron at high-pressure, may be able to determine the degree of supercooling for iron alloys at the high pressure and high temperature, which up-to-now remains unknown.
\par
Recently, Huguet et al.~(2018) \cite{hvhw2018} extrapolated the theory and results from the classical nucleation theory to planetary core conditions. They also investigated plausible causes of deviation from the canonical theory. They revealed that the nucleation barrier is likely to be very high even at the center of the Earth's core. The Earth's inner core nucleation paradox arises from the incompatibility between the slow cooling rate of the Earth's core and the high energetic barrier for homogeneous nucleation. While heterogeneous nucleation seems to be unlikely at the center of the Earth's core \cite{hvhw2018} (it is located thousands of kilometers away from any solid surface), the presence of the solid surface at the core-mantle boundary should ease the nucleation in a top-down scenario of crystallization \cite{hhvj2018}. 

Motivated by this work, several studies have attempted to resolve the paradox using molecular dynamics simulations \cite{wawd2023,wwad2021,szmwh2022,dpa2019}. The presence of light elements can either reduce or increase the nucleation barrier \cite{wawd2023}. In a two-step scenario of nucleation, Sun et al.,~(2022)~\cite{szmwh2022} found that the body-centered cubic (bcc) phase of iron has a lower nucleation barrier than the stable hexagonal close-packed (hcp) phase. However, the nucleation barrier remains significantly high even considering light elements or complex pathways of nucleation. This energetic barrier might be particularly significant for the existence of a slurry layer since the latter requires nucleation sites in the bulk: the origin of these nucleation sites remains a matter of speculation in planetary cores \cite{rl1983}, especially for bottom-up solidification in which no solid phase is initially present \cite{hvhw2018}. The recent molecular dynamic \cite{wawd2023} and geodynamical studies \cite{lkc2020} further highlight the importance of the nucleation process in the crystallization of planetary cores, which remains a subject of active research.

We will now describe in more detail the dynamics of a mushy, and then of a slurry layer in the context of planetary cores. We will highlight fluid dynamics or crystallization experiments that unravel the dynamics of such layers in cores.

\section{Mushy layer}
\label{sec:mushylayer}

Here we present the growth of a mushy layer and its dynamics in the context of planetary cores. Mushy layers have been considerably studied through theoretical, numerical, and experimental approaches because of their importance in industrial casting and sea ice dynamics (see e.g., \cite{bw1995,w1997,fuww2006}). Experiments have used metallic alloys or salt solutions \cite{cgjh1970,sh1988,cc1991,tj1992,wk1994,ww1997,habdll2016} (see Figures \ref{fig:dendritevsequiaxed}b and \ref{fig:mush_height}). Ammonium chloride and sodium chloride solutions have proven good analogs for studying the solidification of metals \cite{jhus1966}. Following these works, dedicated experimental studies have been performed to unravel the effect of convection \cite{bmshca2005,habdll2016} and magnetic field \cite{bfbs1997} on the structure of the Earth's inner core. Note that a core mushy layer differs in terms of structure and evolution compared to the classical casting case: while the dendrites are canonically seen as a steady solid matrix growing from a surface, large planetary dendrites might flow under their weight, melt due to external forces, and sink into the liquid core.

\subsection{State of a mushy layer growing at the Earth's inner core}
\label{subsec:ICmushy} 

Crystallization of the inner core is a major source of energy for convection in the outer core, which generates the Earth's magnetic field. Reciprocally, convection alters the transfer of light elements to the inner core's surface, significantly impacting crystallization. Fearn et al.~(1981) \cite{flr1981} showed that the inner core boundary conditions lead to the formation of dendrites \cite{lr1981}. Based on thermodynamic arguments, they demonstrated that this dendritic mushy layer might extend down to the center of the Earth's core \cite{flr1981}. The mushy zone can be subject to convective instabilities \cite{w1992}: the boundary layer mode and the mushy layer mode. The first mode would be unstable in the inner core since the solute released upon solidification is lighter than the ambient \cite{habdll2016}. The mushy layer mode of convection is controlled by the mushy layer Rayleigh number $R_m$ \cite{wwo2010,wwo2011,wwo2013,rw2013} written as:
\begin{equation}
R_m=\frac{\beta^\star \Delta C g \Pi(\Phi) h }{\nu \kappa},
\label{eq:Rayleigh_R}
\end{equation}
where $\beta^\star = \beta - \Gamma \alpha$ is the expansion coefficient due to temperature and concentration changes ($\beta$ is the solute expansion coefficient, $\alpha$ is the thermal expansion coefficient, $\Gamma$ is the liquidus slope), $\Delta C$ is the difference of composition across the height of the mushy layer, $\nu$ and $\kappa$ are the viscosity and thermal diffusivity of the interstitial liquid. In contrast with the classical Rayleigh number definition, the mushy layer Rayleigh number is proportional to the thickness $h$ of the layer. The permeability $\Pi(\Phi)$ depends on the solid fraction $\Phi$ and is described by the Kozeny-Carman equation \cite{c1956}
\begin{equation}
\Pi(\Phi)=\frac{\lambda_1^2}{4}\frac{(1-\Phi)^3}{k_0 \Phi},
\label{eq:permea}
\end{equation}
where $\lambda_1$ is the primary interdendritic spacing which is set by the morphological instability of the front of solidification \cite{tj1992}. The parameter $k_0$ is a constant (usually taken equal to 5 for directional solidification) but depends on the mean flow direction \cite{tsosa2019}. The mushy layer mode is unstable when the critical Rayleigh number is above $25$ \cite{tj1992}. The effective Rayleigh number is limited by permeability, as vigorous convection induces more crystallization which increases the solid fraction and so decreases the permeability \cite{habdll2016}. This second mode of convection would usually lead to the formation of chimneys as the iron-depleted liquid may dissolve iron dendrites in regions of upwelling \cite{w1992}. The formation of chimneys is well documented by theoretical \cite{w1991,w1997,rw2013,wwo2013} and experimental works in metallurgy \cite{cgjh1970,mh1970,sh1988,bfbs1997}. Based on observations in experiments and evidence from meteorites \cite{eb1982}, Bergman et al.~(1997) \cite{bfbs1997} suggest that the chimneys could be several hundreds of meters large at the inner core boundary. But in the presence of a magnetic field, the convection may take a different form than the chimneys-mode \cite{bf1994}. A strong enough magnetic field might suppress chimney convection \cite{bfb1999}, even if the mushy layer is convectively unstable.
\begin{figure}
\centering
\includegraphics[width=\textwidth]{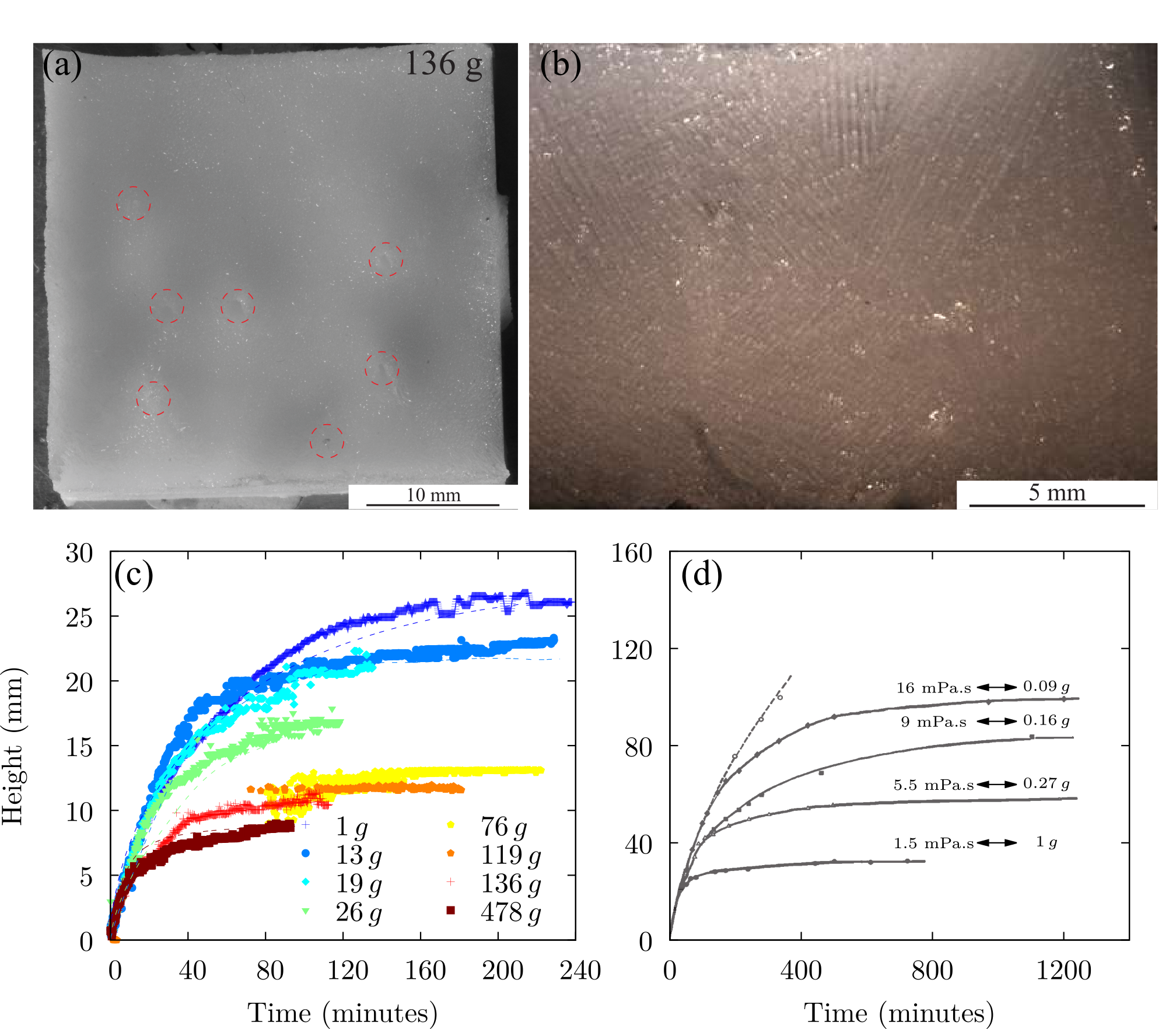}
\caption{(a) Top view of a mushy layer growing under high gravity level (136~g) \cite{h2014,habdll2016}. Red circles denote the chimneys, which are evidence of mushy layer convection. (b) Side view of the mushy layer consisting of ammonium chloride dendrites. Dendrites formed grains with different preferential orientations \cite{h2014}. (c) The thickness of the mushy layer as a function of time for different apparent gravities. Squares are the experimental results and dashed lines are predictions from the evolution model \cite{habdll2016}. (d) Same as (c) for different viscosities, which correspond to different gravity levels given as an indication \cite{tj1992}. Panels (a,b) and (c,d) adapted with permission from Huguet (2014) \cite{h2014} and Huguet et al.~(2016) \cite{habdll2016}, respectively.}
\label{fig:mush_height}
\end{figure}

The particular conditions at the inner core boundary are different from those in the canonical metallurgy literature, due to the simultaneous presence of vigorous convection in the outer core and a very slow inner core growth. Most of the experimental studies \cite{bmshca2005,habdll2016} on the mushy layer at the inner core boundary have explored the effect of the convection in and above the mushy layer on its internal structure (solid fraction, size of the dendrites, or presence of chimneys). To investigate the effect of the vigor of this convection, several experimental studies have been performed by increasing or decreasing the Rayleigh number $R_m$ (see Eq.~\ref{eq:Rayleigh_R}) \cite{tj1992,habdll2016}. Tait and Jaupart (1992) \cite{tj1992} used a viscosifier to decrease the Rayleigh number. Huguet et al.~(2016) \cite{habdll2016} increased the apparent gravity $g$ to increase the Rayleigh number $R_m$. Figure~\ref{fig:mush_height} shows the time evolution of the thickness of a growing mushy layer. Since all other controlling parameters (composition and heat flux) are similar between the experiments, the final thickness of the mushy layer is a proxy of the solid fraction as a function of the Rayleigh number. The solid fraction is significantly increased by the vigorous convection of the liquid in and above it.
Extrapolated to the inner core conditions, Huguet et al.~(2016) \cite{habdll2016} suggested that the inner core mushy layer has a solid fraction very close to one, meaning that the remaining liquid is very sparse.

\begin{figure}
\centering
\includegraphics[width=\textwidth]{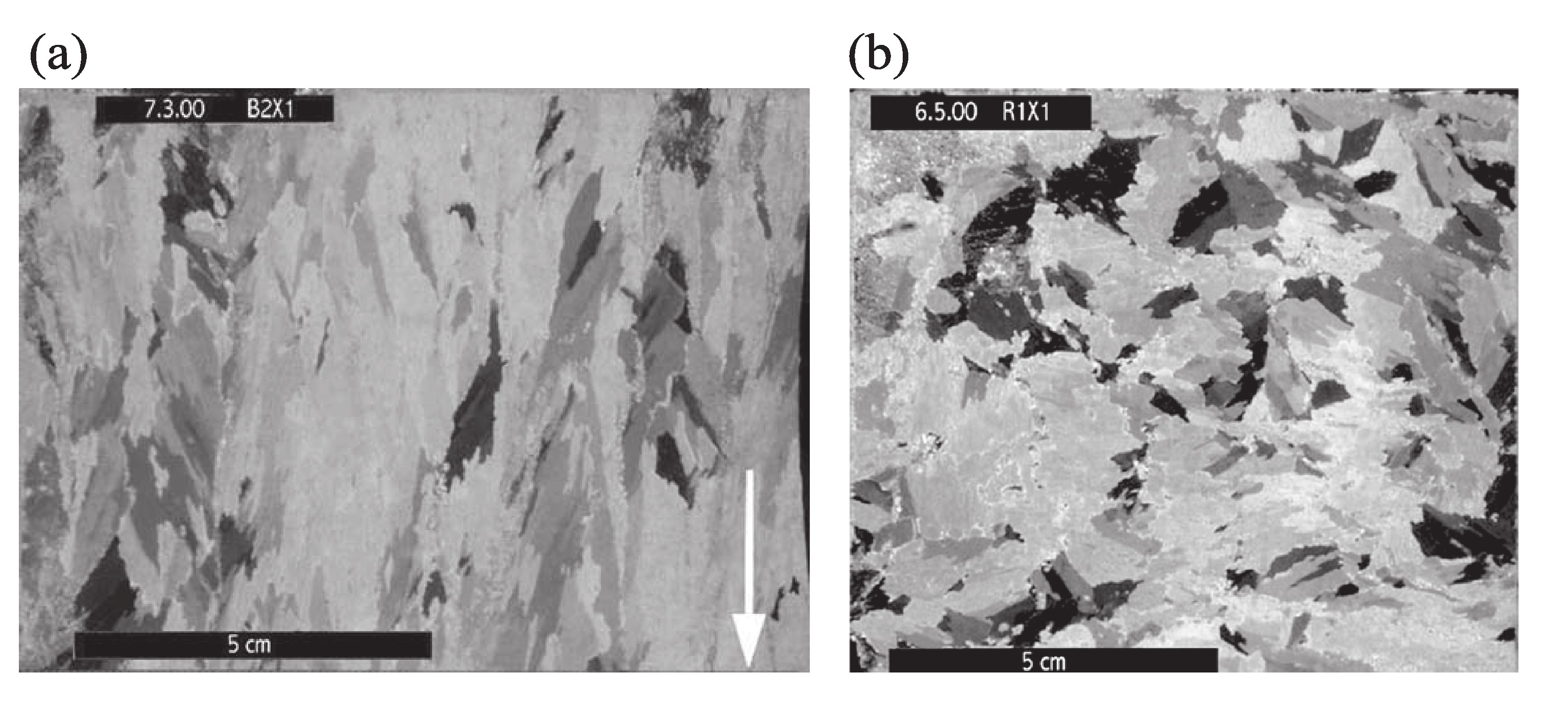}
\caption{Transverse thin section of sea ice crystals with a mean flow parallel to the front of solidification (a) and with no mean flow (b). The white arrow in (a) shows the direction of convective flow near the interface. Panels (a and b) adapted from Bergman et al.~(2002) \cite{bcj2002} with permission of John Wiley and Sons.}
\label{fig:bergman_mush}
\end{figure}

At the inner core boundary, the heat flux extracted by the convection induces a directional solidification of the mushy layer, which produces a preferred orientation of the crystals \cite{hh1962,c1964,kfr2023}. Their orientation is determined by the crystal lattice, which depends on pressure, temperature, and crystal composition. For example,
iron has a body-centered cubic lattice at ambient pressure (1 bar), but at high pressure, the most stable lattice for iron is still debated, between a hexagonal-closed package lattice \cite{thot2010} and body-centered cubic \cite{vwgbdpa2008,blfzds2017}. Analog materials commonly used in mushy layer studies, such as ammonium chloride have a body-centered cubic lattice, and ice growing from salty water has a hexagonal closed package lattice. However, Bergman et al.~(2002) \cite{bcj2002} showed that in directional solidification experiments with hexagonal close-packed crystals, the fluid motions above the mushy layer force the crystals to have a preferred orientation parallel to the flow (Figure \ref{fig:bergman_mush}). With a laboratory model of the inner core growth, Bergman et al.~(2005) \cite{bmshca2005} showed that the convective patterns in the liquid induce a texturing of the mushy layer, which might explain the structure of the inner core as seen by seismological studies \cite{d2014,wipt2023}. 

The typical length scale of the iron crystals (dendrites or equiaxed crystals) is controlled by crystallization at the inner core boundary. The primary spacing between dendrites (related to the grain width) is proportional to the growth rate and the chemical and temperature gradients \cite{kfr2023}. Based on morphological analysis \cite{dab2007} and experimental studies \cite{b1997,b1998}, the typical size of the dendrites in the Earth's core is thought to be between $10^{-1}$ and $10^2$~m. Experiments with high apparent gravity have also shown that the interdendritic spacing and grain size decrease with the vigor of convection \cite{habdll2016}.
Besides, at the length and time scales of laboratory experiments \cite{bmshca2005,bfbs1997,habdll2016}, the network of dendrites is seen as a fixed matrix in the mushy layer. But in planetary cores, dendrites can also compact under their weight on a very long timescale \cite{sykh1996}. Theoretical and numerical works \cite{sykh1996,lkc2020} show that the compaction of a mushy layer will depend on the viscosity of the solid, the mush permeability (related to the solid fraction), and the growth rate of the solid. At the inner core conditions, Shimizu et al.~(1996) \cite{sykh1996} predict that the mushy layer will compact on the length scale of a few kilometers. 

In conclusion, the Earth's inner core mushy layer, if it exists at all, is likely very thin and very depleted in liquid, with strong lateral heterogeneity of crystal orientation as fluid flow and magnetic field induce localized preferred orientation. The presence of an initial supercooled layer, as suggested by the nucleation paradox \cite{hvhw2018}, may imply a phase of rapid growth that changes the structure of the mushy layer. Therefore, the history of the crystallization of the mushy layer is hidden in the layering of the Earth's inner core \cite{lkc2020}.

\subsection{Mushy zone at the top of planetary cores}

\begin{figure}
 \centering
 \includegraphics[width=1\textwidth]{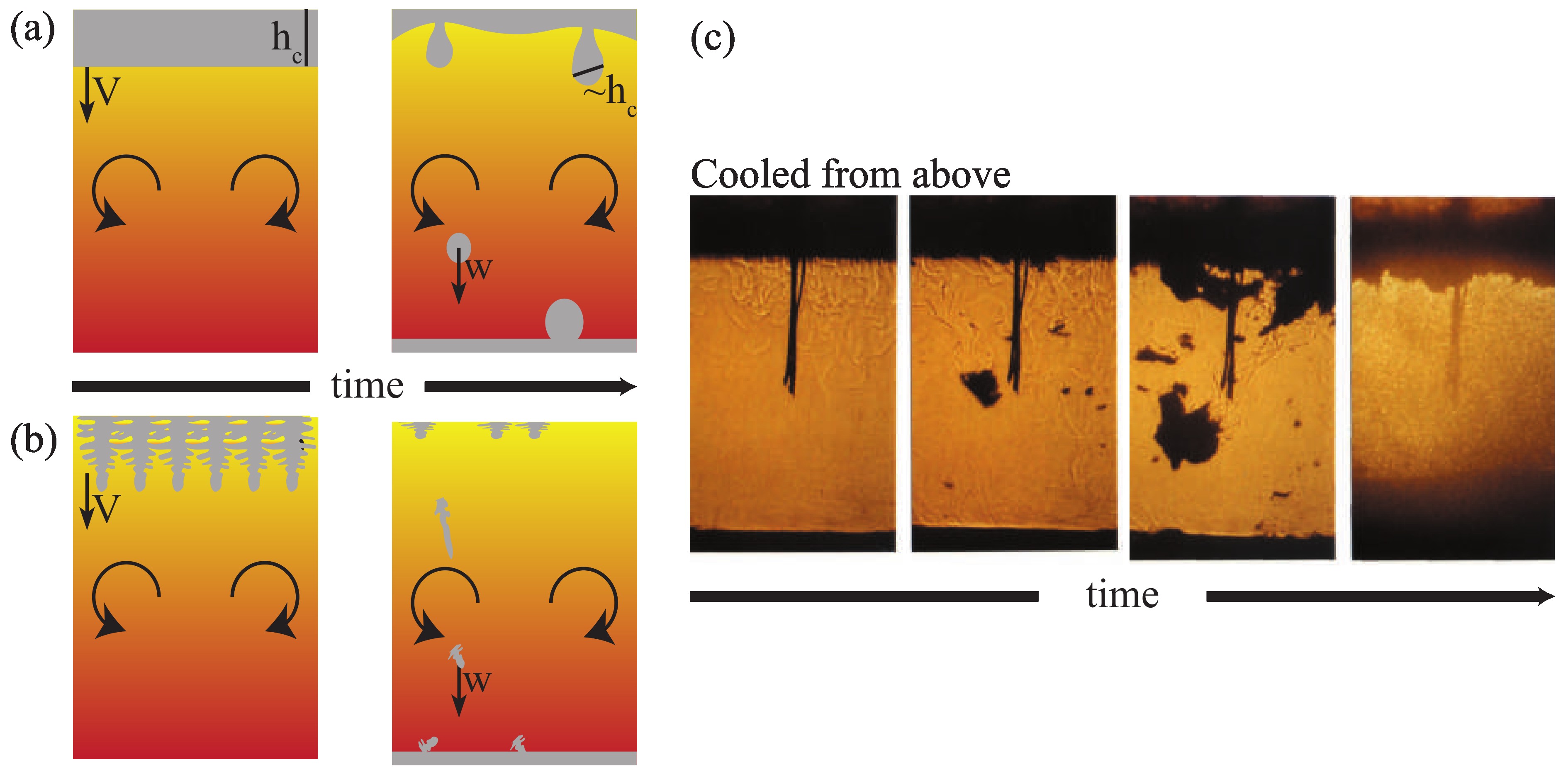}
 \caption{(a,b) Schematic view of the destabilization of a growing mushy layer (with a growth rate $V$) at the top of the core of a small planet \cite{nbn2019,hhvj2018}. (a) Rayleigh-Taylor instability of the iron layer. The wavelength of the instability is about the thickness of the layer $h_c$ \cite{nbn2019}. (b) Necking of independent dendrites. The typical dendrite size depends on the iron's rheology \cite{hhvj2018}. In (a) and (b) iron crystals fall into the liquid core (with a falling velocity $w$) and remelt or accumulate at the bottom. (c) Photos of an experiment using ammonium chloride aqueous solution cooled from above \cite{c2001}. A large chunk of the mushy layer falls into the liquid and produces the crystallization of tiny equiaxed crystals. A temperature probe is inserted into the middle of the tank. Panel (c) adapted with permission from Chen (2001) \cite{c2001}.} 
 \label{fig:topmush}
\end{figure}

In section~\ref{subsec:ICmushy}, we discussed the state of the mush growing from the center of the Earth's core, but as we saw in section \ref{subsec:scenariosSolidification}, crystallization can occur from the core-mantle boundary. While many studies argue in favor of the formation of a slurry layer (snow layer) in this regime of crystallization \cite{had2006,rbs2015} (see section \ref{sec:slurryLayer}), other studies for small planetary cores suggest that iron crystals grow from the core-mantle boundary \cite{ch2006,sesb2016,hhvj2018,nbn2019} (Figure~\ref{fig:topmush}).

Based on the observation of metallic meteorite samples, Haack and Scott~(1992) \cite{hs1992} suggested that dendrite can inwardly grow attached to the core-mantle boundary. Such a mushy layer may differ from the one at the inner core boundary as composition, temperature, and pressure conditions are quite different. The existence of a mushy layer at the top of a core depends on the morphological stability of the interface. Deguen (2009) \cite{d2009} suggests that conditions at the core-mantle boundary of the asteroid might be compatible with an unstable interface and, therefore the formation of a mushy layer.

In a mushy layer composed of iron solid dendrite and a hypo-eutectic liquid, growing from the top will lead to a stably stratified interstitial liquid and unstable dendrites (as they are denser than the surrounding liquid \cite{habdll2016,nbn2019}). The absence of convection and the slow growth rate in the mush induce a highly porous mushy layer, in contrast to the highly compact mushy layer at the inner core boundary \cite{habdll2016}. Therefore, the mushy layer can be seen as a solid interconnected matrix layer with viscosity depending on the porosity \cite{nbn2019} (Figure~\ref{fig:topmush}a), or as independent dendrites growing towards the center \cite{hhvj2018} (Figure~\ref{fig:topmush}b). A mushy layer will be subjected to Rayleigh-Taylor instability \cite{nbn2019}. The length and time scales of the delamination of the mushy layer depend on the rheology and the solid fraction, which are controlled by the strain, rate, temperature, and composition.

Experimental studies of unstable mushy layers \cite{c2001,ante2020} 
have shown that large chunks of a mushy layer, with a typical size ranging between the interdendritic spacing and the mush thickness, can fall into the liquid (Figure~\ref{fig:topmush}c). However, the rheology of salt or ice dendrites is in the brittle regime. The fragmentation of dendrites is often observed in experiments due to the vigorous convection and strong strain rate applied to secondary or tertiary arms of dendrites \cite{jhus1966,ccr2004,skwl2018}. In planetary cores, the rheology of iron at high pressure and temperature is still poorly known \cite{rt2020} and the strain rate is likely to be small (for example, the typical convective velocity for the Earth's core is of the order of mm/s).

On another end, independent dendrites attached to the core-mantle boundary can be seen as slab necking under gravity during the subduction process in the Earth's mantle \cite{rfbm2008,dsg2012,bsr2015}. The necking and delamination of dendrites could produce a large range of crystal sizes falling into the liquid core \cite{hhvj2018,nbn2019}. Moreover, the fragmentation of dendrites induces the formation of equiaxed crystals through the production of secondary nucleation sites \cite{jhus1966,ccr2004}. It is also well-known that stronger convection will produce more fragments resulting in a larger amount of equiaxed crystals in the bulk \cite{skwl2018}. 

Laboratory experiments and theoretical studies applied to the growth of the top-down mushy layer are missing to determine which destabilization mode dominates. One might suggest that the existence of an iron snow might result from the growth of a thin mushy layer producing a high quantity of small crystals falling into the liquid cores, as further described in section \ref{sec:slurryLayer}. The existence of such a regime would solve the paradoxical existence of a large nucleation barrier for homogeneous nucleation since primary dendrites would grow from the core-mantle boundary and secondary nucleation would provide sites for the iron snow \cite{d1984}.

\subsection{Melting of a mushy layer at the inner core boundary}

\begin{figure}
 \centering
 \includegraphics[width=1\textwidth]{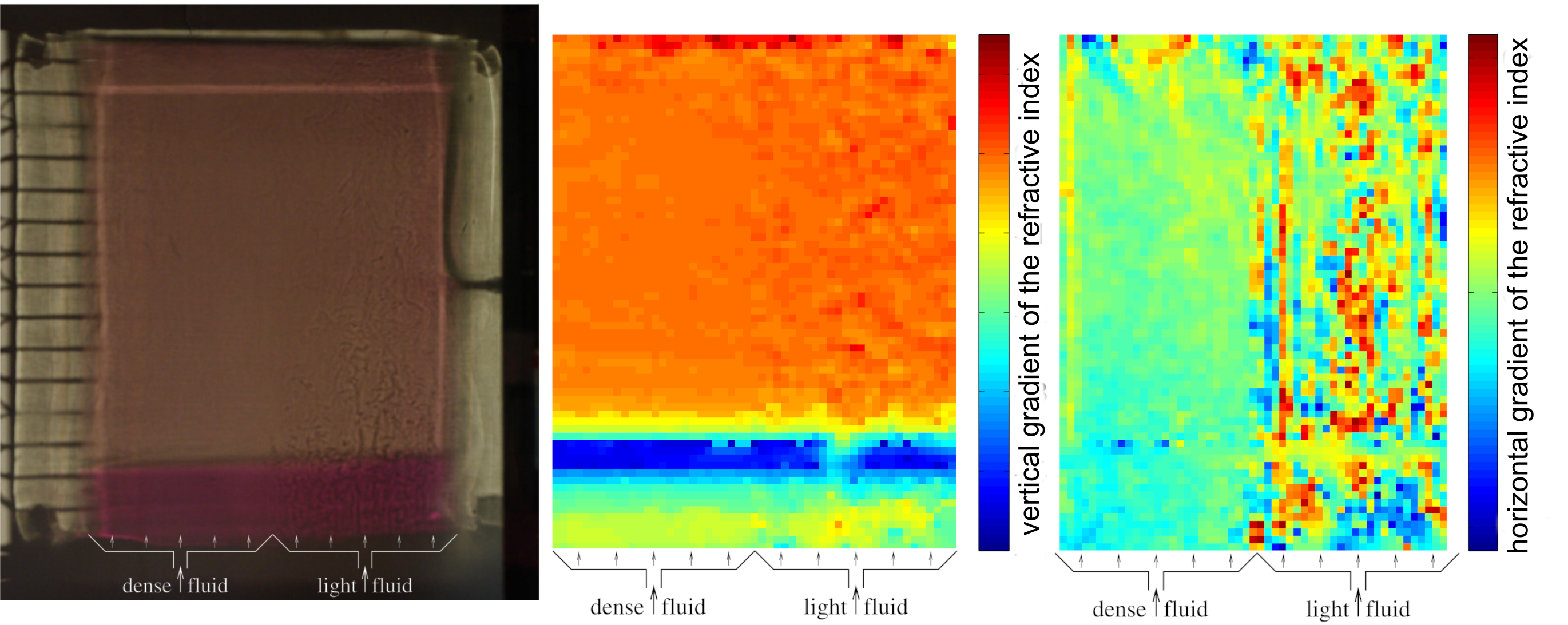}
 \caption{Convection experiment with light and dense fluid released from the bottom and formation of an F-layer \cite{adm2010}. The shadowgraph on the left shows the dense fluid (with purple dye) and light fluid rising through. Using a synthetic Schlieren method, the vertical gradient (in the middle) and the horizontal gradient (on the right) of the refractive index reveal the dense layer and the convective plume, respectively. Figure adapted with permission from Alboussière et al.~(2010) \cite{adm2010}.}
 \label{fig:FlayerConv}
\end{figure}

Finally, Earth's seismology has highlighted the existence of the transition region between the outer core and the inner core \cite{sp1991,ok2015}. This region, the so-called F-layer, is characterized by the observation of slower P-waves than predicted by the Preliminary Reference Earth Model (PREM) \cite{da1981}, which assumes a well-mixed outer core. Therefore, the F-layer is expected to be a denser and stably stratified fluid layer. Its origin is still debated but one explanation comes from the melting of the inner core. Melting of the inner core would release a denser liquid enriched in iron, sitting at the base of the outer core. Several studies have examined mechanisms for such melting including a translation convection mode of the inner core \cite{adm2010,mcms2010} and a heterogeneous heat flux at the inner core boundary \cite{gsmr2011}. A model experiment was set up in Alboussière et al.~(2010) \cite{adm2010} to mimic the phase changes induced by the translation of the inner core (Figure~\ref{fig:FlayerConv}). One hemisphere crystallizes and releases light elements, while the other melts and releases dense liquid iron with few light elements. This was modeled in a volume of salted water, injecting fresh (light) water on the right side of the bottom surface through a porous surface, while more salted water was injected on the left side. Overall, the buoyancy flux of light elements exceeded slightly that of heavy elements to account for the slow net crystallization of the inner core. Then, a dense layer started to grow from the bottom, providing a mechanism for the existence of the F-layer. Light plumes could still cross that dense layer and drive a convective flow throughout the whole volume. The corresponding flow necessarily exists in the Earth's core, at least to drive the geodynamo. 

\begin{figure}
 \centering
 \includegraphics[width=1\textwidth]{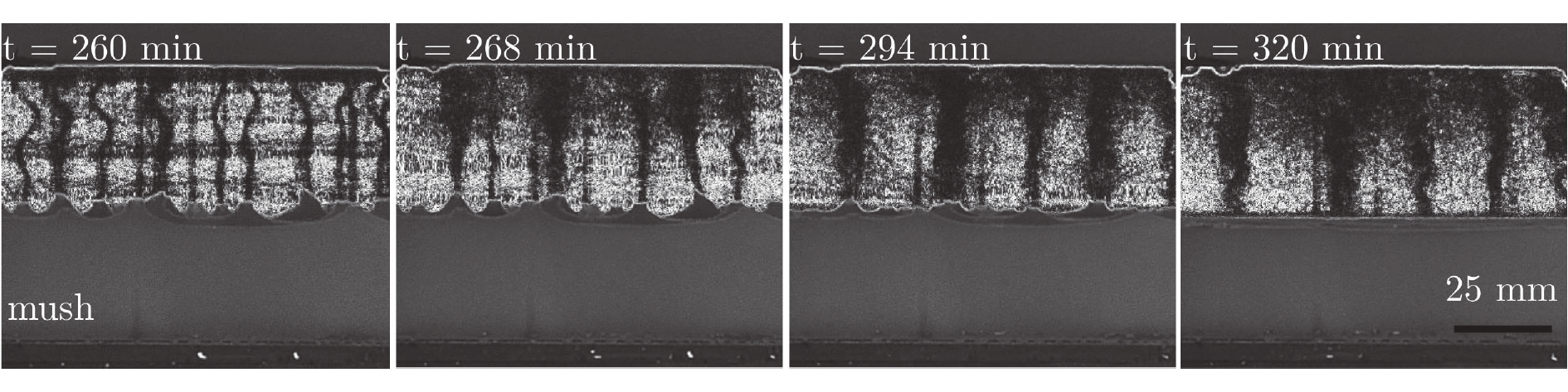}
 \caption{Heating and melting from above of a growing mushy layer \cite{habdll2016}. The crystallization and convection inside the mushy layer induce the formation of chimneys and meandering plumes ($t= 260$~min). Note that double-diffusive convection occurs in the liquid layer as seen in other experiments \cite{c1997}. Once heating and melting start, chimneys melt and mushy layer convection becomes stronger, as evidenced by the turbulent plumes (t=$268,\, 294,\, 320$~min). Figure adapted with permission from Huguet et al.~(2016) \cite{habdll2016}.}
 \label{fig:melting_mush}
\end{figure}

 Several experiments using a reactive porous medium or a mushy layer have shown that the melting implies a change in the structure of the matrix and the liquid composition \cite{hhw2004,hhw2005,ybha2015,habdll2016}. Moreover, melting induces an enhanced crystallization in a deeper part of the mushy layer \cite{ybha2015,habdll2016} as heavy melt drives stronger compositional convection and recrystallization (Figure \ref{fig:melting_mush}). This leads to a higher solid fraction inside the mush \cite{ybha2015,habdll2016}, the formation of a dense layer above the mush, and the disappearance of cones of chimneys \cite{habdll2016} (Figure \ref{fig:melting_mush}).

\section{Slurry layer}
\label{sec:slurryLayer}
In section \ref{sec:cry_planet}, we show that the crystallization of a planetary core can lead to the formation of free crystals in a slurry layer. Here, we examine the fluid dynamics and the phase change of crystals in such a particle-laden environment. A key ingredient of the slurry dynamics is the difference of density between the crystals and the interstitial fluid. When crystals are sufficiently concentrated, this density difference manifests as an effective buoyancy on the ambient fluid, through which crystals may collectively force a large-scale flow beyond the scale of a single crystal. Despite the existence of studies on such flows in other geophysical contexts (landslides \cite{fmy2009}, turbidity currents \cite{nhkm2002,oms2019}, snow avalanches \cite{kms2018}), the planetary situation of crystals settling in a bulk liquid is much different and fewer experiments have been performed to model this phenomenon. In addition, crystal formation and remelting are two complex ingredients that must be included to model core evolution. With an incremental approach, the next subsections first describe the collective dynamics of inert particles (subsection \ref{subsec:collective}) before adding the additional complexity of phase change (subsection \ref{subsec:phasechange_slurry}).

\subsection{Collective motion of inert particles}
\label{subsec:collective}
\subsubsection{What numbers control the motion of particles?}
Suppose a suspension of particles settles in an infinite domain. The motion of a single particle depends on its radius $r_p$, the local gravity $g$, the particle density $\rho_p$, and finally the fluid density $\rho_a$ and kinematic viscosity $\nu_a$. The motion of multiple particles depends on their local volume fraction $\phi$, and if the ambient fluid is moving, it also depends on the ambient fluid velocity $U_a$. According to the Vaschy-Buckingham theorem, four dimensionless numbers can be used to characterize the motion of a suspension in a moving ambient: the density ratio $\Pi_\rho \equiv \rho_p/\rho_a$ quantifies the particles' inertia; the volume fraction $\phi$ controls the collective behavior of particles and their forcing on the fluid; the particle Reynolds number $Re_p \equiv 2 w_s r_p / \nu_a$ characterizes the regime of the flow past a single particle. It depends on the terminal settling velocity $w_s$ of a single particle, which is usually implicitly defined as a function of $Re_p$. A classical expression for $Re_p<800$ is \cite{csst2011}:
\begin{equation}
\label{eq:wsSchillerNaumann}
w_s = \frac{w_s^{\textrm{Stokes}}}{1+0.15Re_p^{0.687}},
\end{equation}
where the Stokes velocity is defined as
\begin{equation}
\label{eq:wsStokes}
w_s^{\textrm{Stokes}} = \frac{2gr_p^2(\rho_p-\rho_a)}{9\nu_a \rho_a}.
\end{equation}
Finally, the Rouse number $\mathcal{R} \equiv w_s/U_a$ quantifies how fast particles gravitationally drift with respect to ambient fluid motions. This can be shown from the momentum equation of a single small particle settling in a moving fluid: for a sufficiently small particle with low inertia (see \cite{kfl2023} for details), the buoyancy of this particle balances its drag that is proportional to the difference between the local fluid velocity $\bm{v}_a$ and the local particle velocity $\bm{v}_p$. This readily gives $\bm{v}_p = \bm{v}_a - w_s \bm{e}_r$ where $\bm{e}_r$ is the radial unit vector pointing outward from the core center. Such a small particle follows the ambient flow and constantly drifts in the direction of gravity with a velocity $w_s$; the Rouse number quantifies the relative gravitational drift between the particle and the local fluid motions with $|\bm{v}_p - \bm{v}_a|/|\bm{v}_a| = w_s/|\bm{v}_a|=\mathcal{R}$. This 4D parameter space is a reference to characterise the dynamics of inert monodisperse particles with a locally uniform volume volume fraction $\phi$. Additional ingredients of relevance for planetary interiors (e.g. heterogeneities of concentration, phase change, polydispersity...), that would largely expand the size of this parameter space, are incrementally considered in the next sections.
\par
The difficulty in resolving the particulate dynamics in large-scale turbulent geophysical flows has led many experimentalists to model slurries as a fluid whose effective density (and viscosity) differs from the ambient -- for example using salt water instead of a dense suspension of particles. Such an approximation corresponds to the limits $Re_p \rightarrow 0$ and $\mathcal{R} \rightarrow 0$. We show below some experimental works that revealed the need to refine the description of the particles' dynamics.\\

To gain understanding on the dynamics of settling-driven convection, two different approaches can be found: some authors focus on the fundamental building blocks that nourish this convection at a local scale -- continuous releases of buoyancy in \textit{particle-laden plumes}, and instantaneous releases of buoyancy in \textit{particle clouds} or particle-laden thermals -- while other authors investigate the slurry dynamics directly as a whole with a global approach. Numerical simulations and models of core evolution usually adopt this second vision and consider slurry layers like the snow zone as a uniform region with quasi-steady dynamics (e.g., \cite{rbs2015}). Yet, section \ref{subsec:nucleation} suggests that the slurry dynamics is neither steady nor uniform in the snow zone. Therefore, the motions produced in the snow zone might be described as a sum of localized releases of snowflakes in different locations and at different times. How should these releases be parameterized in terms of crystal sizes, concentrations, distance in space, and delay in time between two bursts of crystals? Should these releases be instantaneous or have a finite duration? In this regard, studies of local convective structures bring some light on the dynamics that might be at play in slurry layers. Both the local and global approaches are addressed respectively in sections \ref{subsubsec:localConvection} and \ref{subsubsec:globalConvection}. Since we are interested in large-scale flows in liquid iron that has a low viscosity $10^{-6} \unit{m^2.s^{-1}}$, these flows are likely {turbulent}, which is why we restrict the description to such flows.

\subsubsection{\label{subsubsec:localConvection}Modelling convection with a local approach}

\begin{figure}
\centering
\begin{subfigure}[b]{.46\textwidth}
\centering
\caption{}
\includegraphics[width=\textwidth]{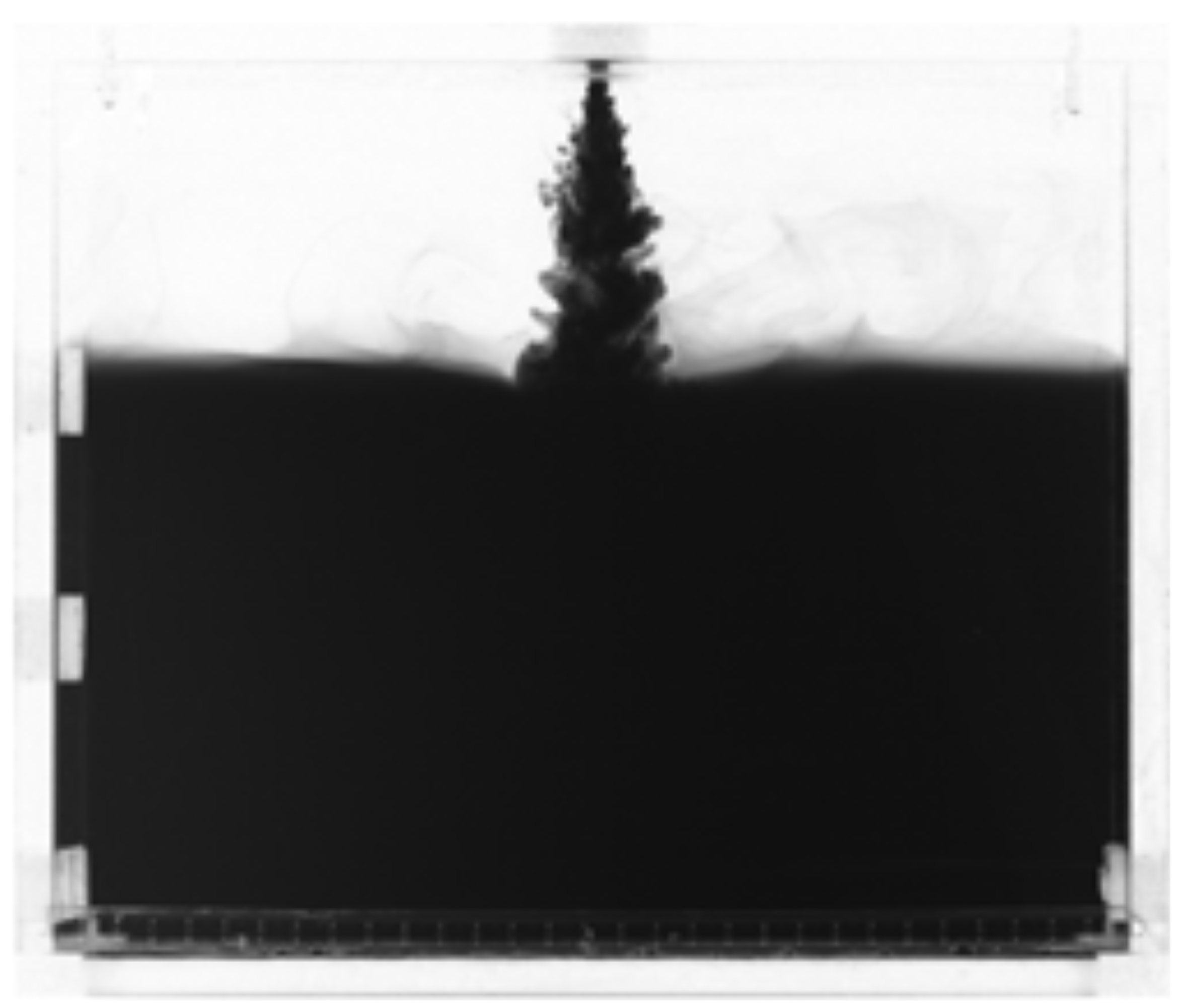}
\label{subfig:fillingBox}
\end{subfigure}
\begin{subfigure}[b]{.4\textwidth}
\centering
\caption{}
\includegraphics[width=\textwidth]{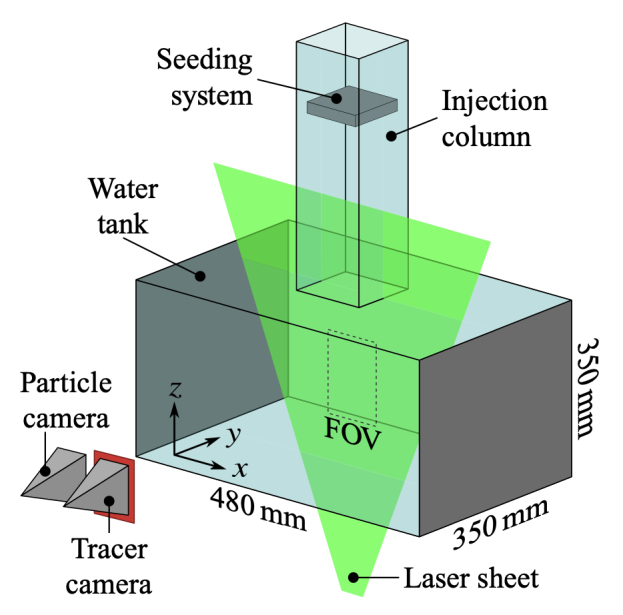}
\label{subfig:setupZurner2023}
\end{subfigure}
\caption{(a) Steady-state reached in a `filling box' experiment with a dyed salty plume generated in fresh water. (b) Experimental setup used by Zürner et al.~(2023) \cite{ztdmm2023} to generate a central settling-driven plume from the injection column into the water tank. Panels (a) adapted from Dadonau et al.~(2020) \cite{dpl2020}, (b) adapted from  Zürner et al.~(2023) \cite{ztdmm2023}, with permission of the Cambridge University Press and the American Physical Society, respectively.}
\label{fig:particleLadenPlumes}
\end{figure}

Particle-laden plumes are often studied in a steady state with the `filling box' technique (see the illustration with a one-phase plume in Figure~\ref{subfig:fillingBox}): from an inlet at the center of the top wall of the tank, a first source feeds a negatively buoyant plume with a volume flux $Q_p$; in the meantime, a second source feeds a volume flux $Q_\mathrm{in}$ that nourishes a downward displacement flow in the whole tank; finally, a sink lets a volume flux $Q_\mathrm{out}$ spill from the tank, thus maintaining a constant volume of fluid in the tank. This setup is particularly relevant to quantify the plume entrainment which characterizes its coupling with the ambient fluid. McConnochie et al.~(2021) \cite{mcm2021} recently used this technique with particle plumes. They revealed that particles modify the entrainment rate of such steady plumes when they settle in opposite directions to the initial volume flux, e.g., when downward-settling particles rain out from an upward-oriented plume. Other experiments by Zurner et al.~(2023) \cite{ztdmm2023} focused more precisely on the coupling between the particulate motion and the flow. In their experiments, particles were continuously sieved from the top of an injection column and formed a steady settling-driven plume -- see Figure~\ref{subfig:setupZurner2023}. By measuring both the local fluid velocities through PIV and the velocity of each particle through PTV, the authors quantified the energy transfer from the particles to the fluid. Their results show that particles of low particle Reynolds number are more efficient at transferring energy to the flow than particles of large particle Reynolds number. This was interpreted as a consequence of the difference in near-field flow around particles.
\par
In both experiments \cite{mcm2021,ztdmm2023}, the particulate nature of the buoyancy forcing directly affects the particle plume, modifying its coupling with the ambient and, if multiple plumes are driving large-scale convection, with other particle plumes. It points to the importance of the density and size of crystals in solidifying cores.
\begin{figure}
\centering
\begin{minipage}{.6\textwidth}
\begin{subfigure}[b]{\textwidth}
\centering
\caption{}
\includegraphics[width=\textwidth]{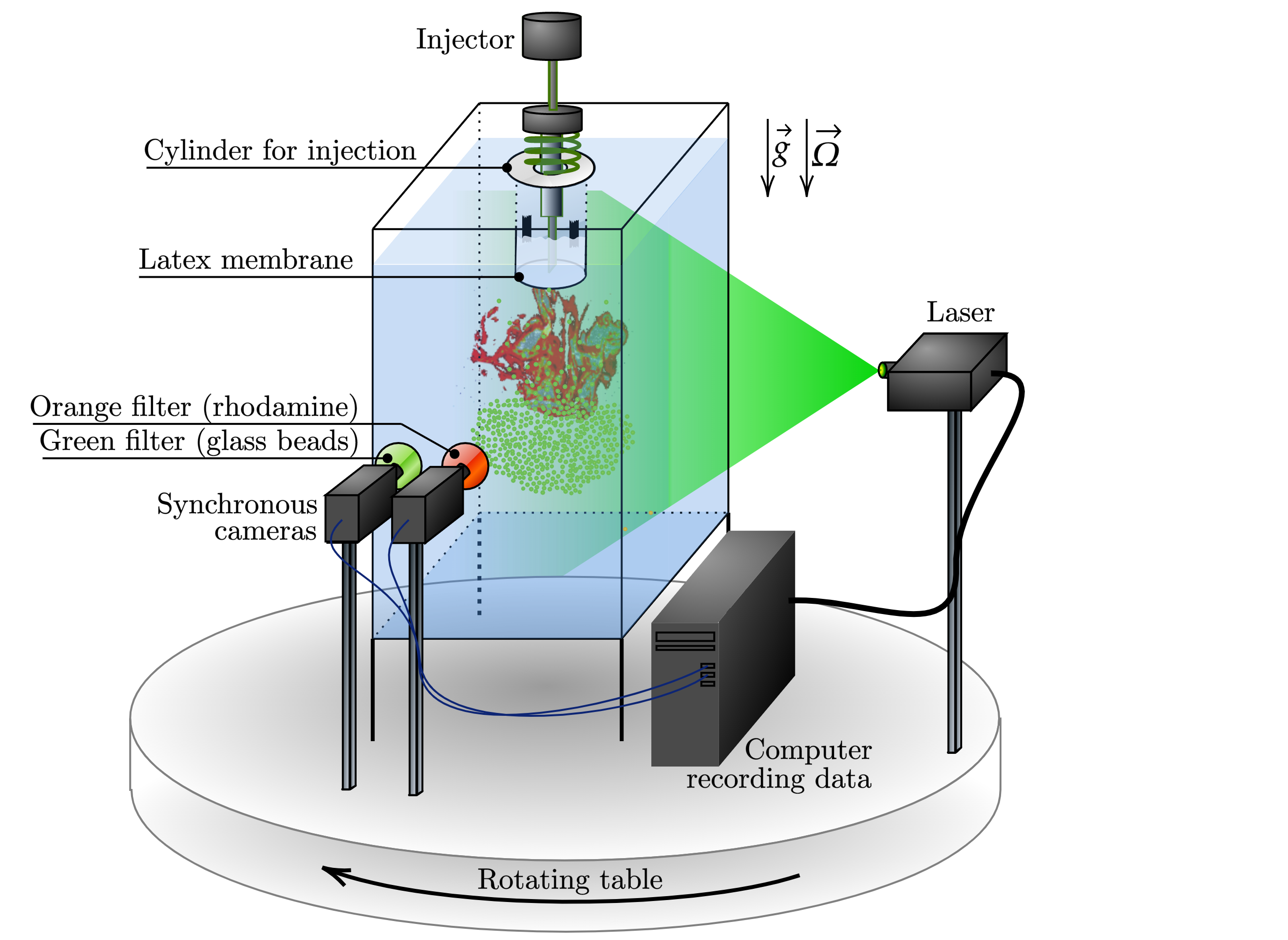}
\label{subfig:setupParticleClouds}
\end{subfigure}
\end{minipage}
\begin{minipage}{.395\textwidth}
\begin{subfigure}[b]{\textwidth}
\caption{$\mathcal{R}=0.308$}
\centering
\includegraphics[width=\textwidth]{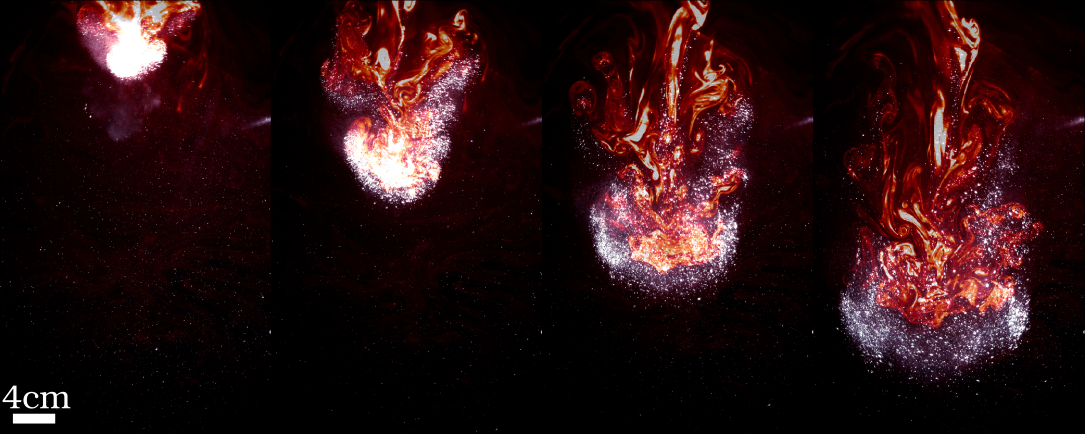}
\label{subfig:montageRhoda90150}
\end{subfigure}
\par
\begin{subfigure}[b]{\textwidth}
\caption{$\mathcal{R}=1.48$}
\centering
\includegraphics[width=\textwidth]{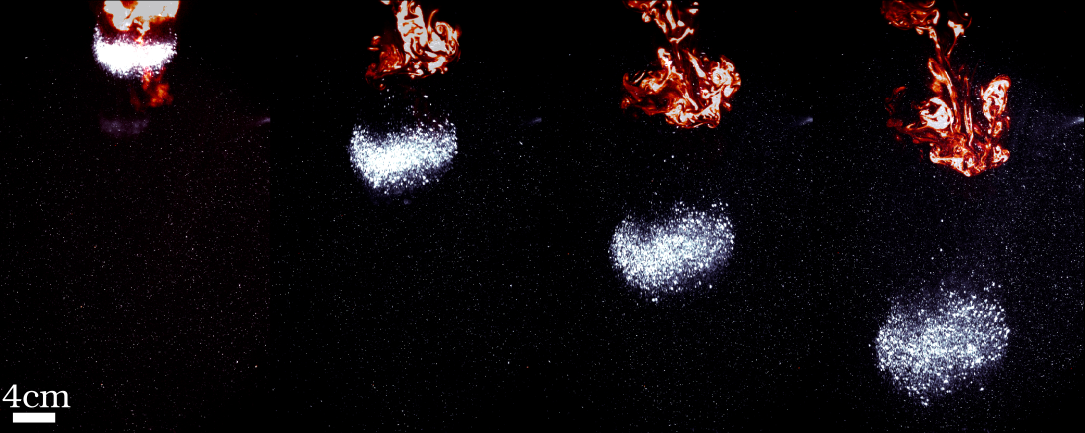}
\label{subfig:montageRhoda250500}
\end{subfigure}
\end{minipage}
\caption{(a) Experimental setup used in Kriaa et al.~(2022) \cite{ksfl2022}. A cylinder partly immersed below the free surface is initially sealed by a latex membrane; after rupturing the membrane with a needle, glass spheres settle in fresh water, reflect the green light from the vertical laser sheet, and their motion is recorded by the camera with the green filter. The orange coloring (rhodamine B) is initially present with particles in the cylinder; as particles fall, this dye is tracked by the camera with an orange filter. The influence of background rotation can be studied since the setup is entirely mounted on a rotating table. (b,c) Snapshots show, in the absence of background rotation, the gradual separation between particles (in white) and the released dye (in orange), all the faster as the Rouse number $\mathcal{R}$ increases from (b) to (c). Time intervals $\Delta t$ between snapshots are (b) $\Delta t=3.0s$ and (c) $\Delta t=1.2s$. Panels (a) and (b) adapted from Kriaa et al.~(2022) \cite{ksfl2022} with the permission of the American Physical Society.}
\label{fig:MontagesRhoda}
\end{figure}

The transient settling of particle clouds reveals some additional physics that depends on the cloud Reynolds number based on its radius and bulk velocity. Several studies have focused on clouds of large cloud Reynolds number (e.g., \cite{rahimipourDynamicBehaviourParticle1992,ruggaberDynamicsSedimentClouds2000,bushParticleCloudsHomogeneous2003,zhaoEffectAirRelease2012,laiTwophaseModelingSediment2013,landeauExperimentsFragmentationBuoyant2014,penasBubbleladenThermalsSupersaturated2021,ksfl2022}). They are typically generated in the lab by releasing some buoyant material from the inlet of an injector that is located above water \cite{zhaoEffectAirRelease2012}, exactly at the free surface \cite{ruggaberDynamicsSedimentClouds2000,bushParticleCloudsHomogeneous2003,landeauExperimentsFragmentationBuoyant2014} or submerged \cite{ruggaberDynamicsSedimentClouds2000,laiTwophaseModelingSediment2013,landeauExperimentsFragmentationBuoyant2014,ksfl2022}. In most experiments, the injector is initially sealed by a membrane that keeps the buoyant material until it is ruptured (see Figure~\ref{subfig:setupParticleClouds}). After rupturing the membrane, the buoyant mixture rolls up and a large-scale turbulent cloud forms in which the particles swirl as long as the cloud falls faster than the settling speed of an individual crystal (see snapshots one and two in Figure~\ref{subfig:montageRhoda90150}); as the cloud dilutes, it decelerates and particles rain out of the cloud when the latter is as slow as their settling speed \cite{deguenExperimentsTurbulentMetalsilicate2011} (see the third snapshot in Figure~\ref{subfig:montageRhoda90150}); after this separation, particles settle almost vertically with a constant velocity (see the fourth snapshot in Figure~\ref{subfig:montageRhoda90150}). The larger the Rouse number, the earlier the separation (compare Figures~\ref{subfig:montageRhoda90150} and \ref{subfig:montageRhoda250500}). It means that a given buoyancy anomaly released from a fixed initial cloud size gradually loses its ability to produce large-scale fluid motions as the particle size is increased, due to the decoupling between eddies and the particles that gravitationally drift with respect to the flow -- in other words, the flow experiences transitions that are bound to the particulate nature of the buoyant material. The entrainment of particle clouds is also affected by the particles' properties: in \cite{ksfl2022} the authors showed that particle clouds have a maximum growth rate and hence dilute the fastest for a finite particle size, 75\% faster than first-order estimates that disregard the particulate nature of the buoyancy forcing.\\

To sum up, these experiments with plumes and clouds show that the particle size, through $Re_p$, $\mathcal{R}$ and $\phi$, affects the efficiency of the collective drag, the decoupling of particles from fluid motions, and the entrainment capacity of these local flows. Therefore, it controls the ability of the slurry to generate a large-scale flow, with consequences for dynamo generation. The next section confirms this conclusion and addresses another key question: When no length scale nor velocity scale is imposed by plumes and particle clouds, how does a slurry layer evolve and what is its spatial structure?

\subsubsection{\label{subsubsec:globalConvection}A global approach of convection}

\begin{figure}
\centering
\includegraphics[width=.9\textwidth]{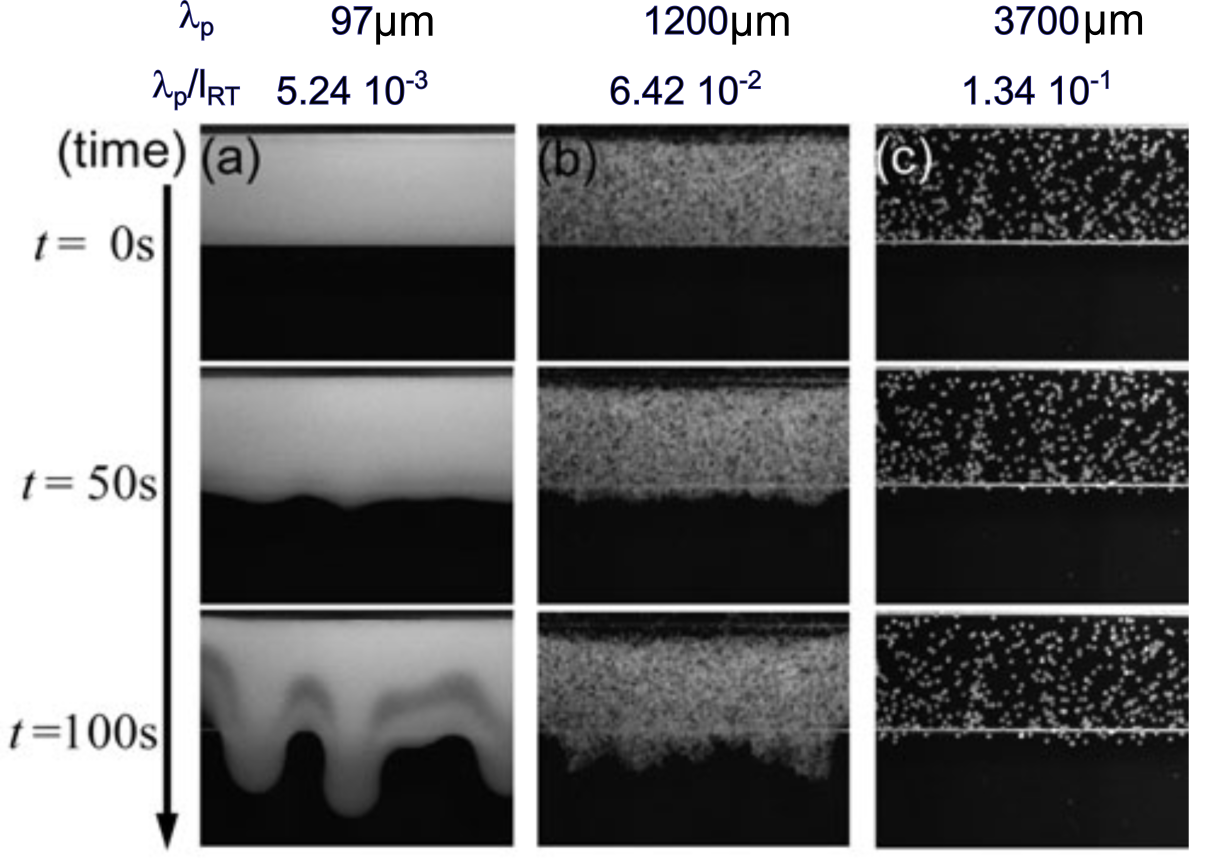}
\caption{Snapshots of experiments where a suspension of white particles settles above the clear fluid in a narrow cell. From left to right the inter-particle distance $\lambda_p$ increases (it is respectively equal to $97 \unit{\mu m}$, $1200 \unit{\mu m}$ and $3700 \unit{\mu m}$), leading to a transition from a collective fluid-like behavior (particles undergo a Rayleigh-Taylor-like fluid instability) to individual particle-like behavior (particles essentially settle vertically). This transition to individual behavior is reflected by the ratio of the inter-particle distance $\lambda_p$ over the expected width of the Rayleigh-Taylor mushrooms $l_\mathrm{RT}$, which equals $5.24\times 10^{-3}$, $6.42\times 10^{-2}$ and $1.34\times 10^{-1}$ from left to right. Figure extracted from Harada et al.~(2012) \cite{haradaParticlelikeFluidlikeSettling2012} with permission from Springer Nature.}
\label{fig:harada}
\end{figure}

Besides the local analysis of convection through its building blocks, some experiments directly analyze the behavior of a whole layer of particles settling in an initially quiescent fluid. The simplest approach is to consider this layer uniform, settling above another layer of clear lighter fluid. This configuration has been investigated in experiments to analyze the onset of the fluid motions at the interface between the two layers through a Rayleigh-Taylor-like instability \cite{boffettaIncompressibleRayleighTaylor2017,careyInfluenceConvectiveSedimentation1997,friesInfluenceParticleConcentration2021}, typically after sliding a gate open. For example, Fries et al.~(2021) \cite{friesInfluenceParticleConcentration2021} let a mixture of water and particles settle above slightly denser sugary water, with a motivation to study the settling and deposition of ash from warm volcanic clouds. Other authors let dense particles settle in the air down a settling column until they penetrate in a tank of constant \cite{mizukamiParticlegeneratedTurbulenceHomogeneous1992,careyInfluenceConvectiveSedimentation1997} or piecewise linear \cite{manvilleVerticalDensityCurrents2004} density profile, and they analyze the particle instabilities that develop near the free surface. Some studies in a Hele-Shaw cell (see Figure~\ref{fig:harada} extracted from Harada et al.~(2012) \cite{haradaParticlelikeFluidlikeSettling2012}) or in numerical simulations \cite{yamamotoNumericalSimulationConcentration2015} have shown that particles behave collectively when the interparticle distance of a uniform suspension, proportional to $r_p \phi^{-1/3}$, is sufficiently small for the hydrodynamic perturbations of neighboring particles to overlap, enabling them to collectively drag the interstitial fluid. A paramount consequence is that the collective settling of particles considerably accelerates their fall compared to the individual settling velocity of isolated particles.

\begin{figure}
\centering
\begin{subfigure}[t]{.49\textwidth}
\centering
\caption{}
\includegraphics[width=\textwidth]{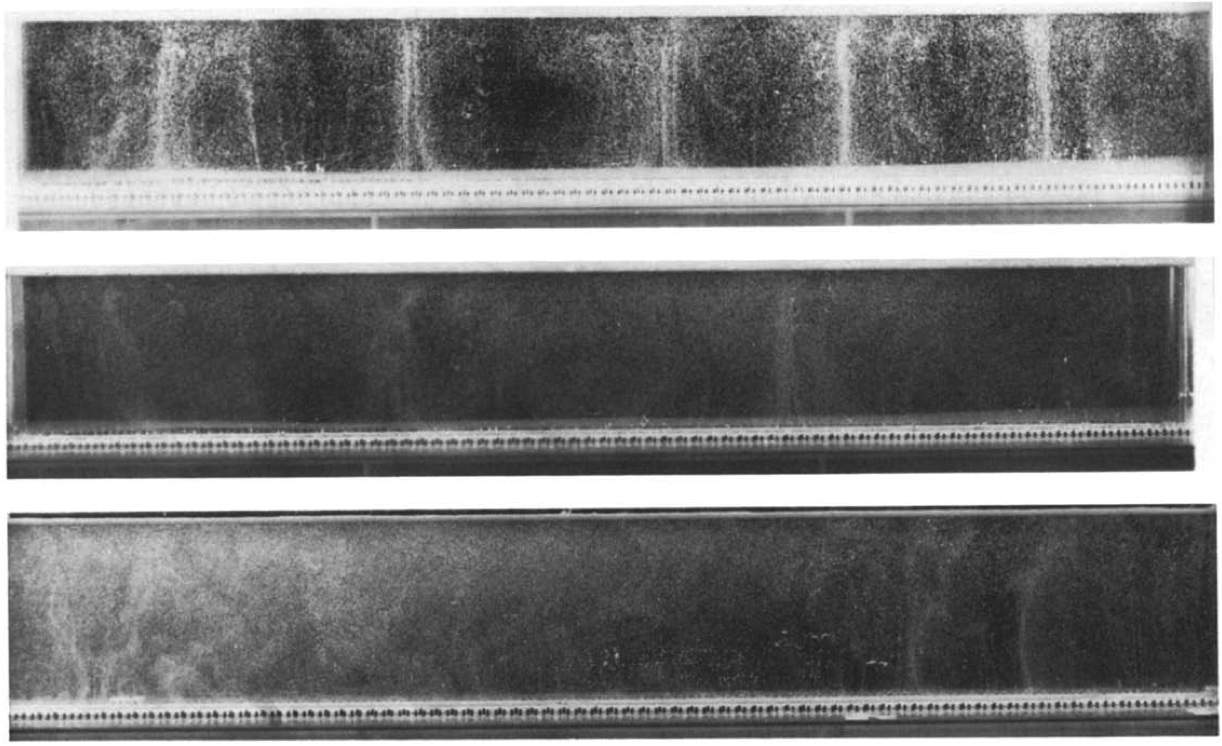}
\label{subfig:KimuraFigure4}
\end{subfigure}
\begin{subfigure}[t]{.49\textwidth}
\caption{}
\centering
\includegraphics[width=\textwidth]{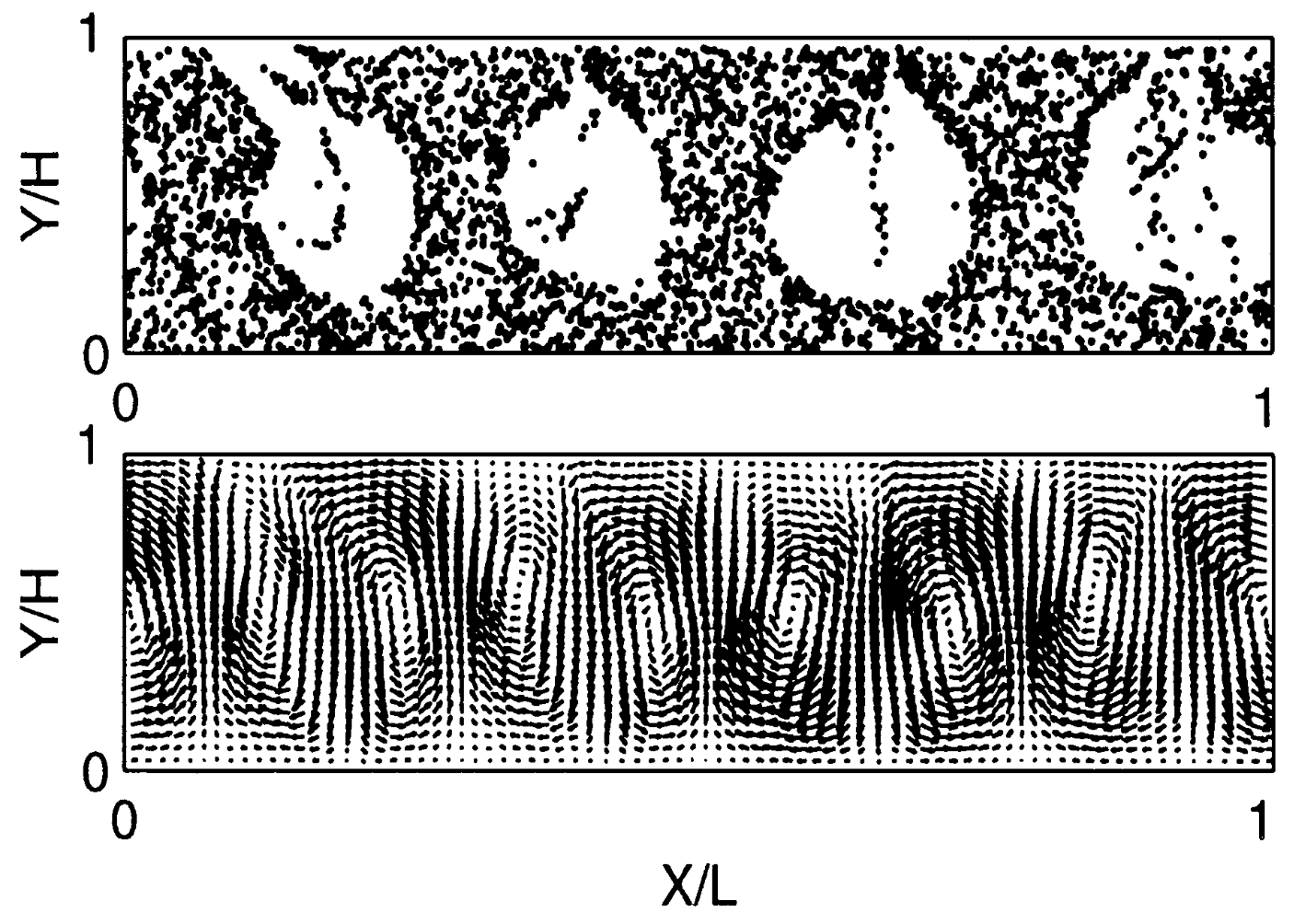}
\label{subfig:ClimentMagnaudetFigure3}
\end{subfigure}
\par
\begin{subfigure}[b]{.49\textwidth}
\caption{}
\centering
\includegraphics[height = 6cm]{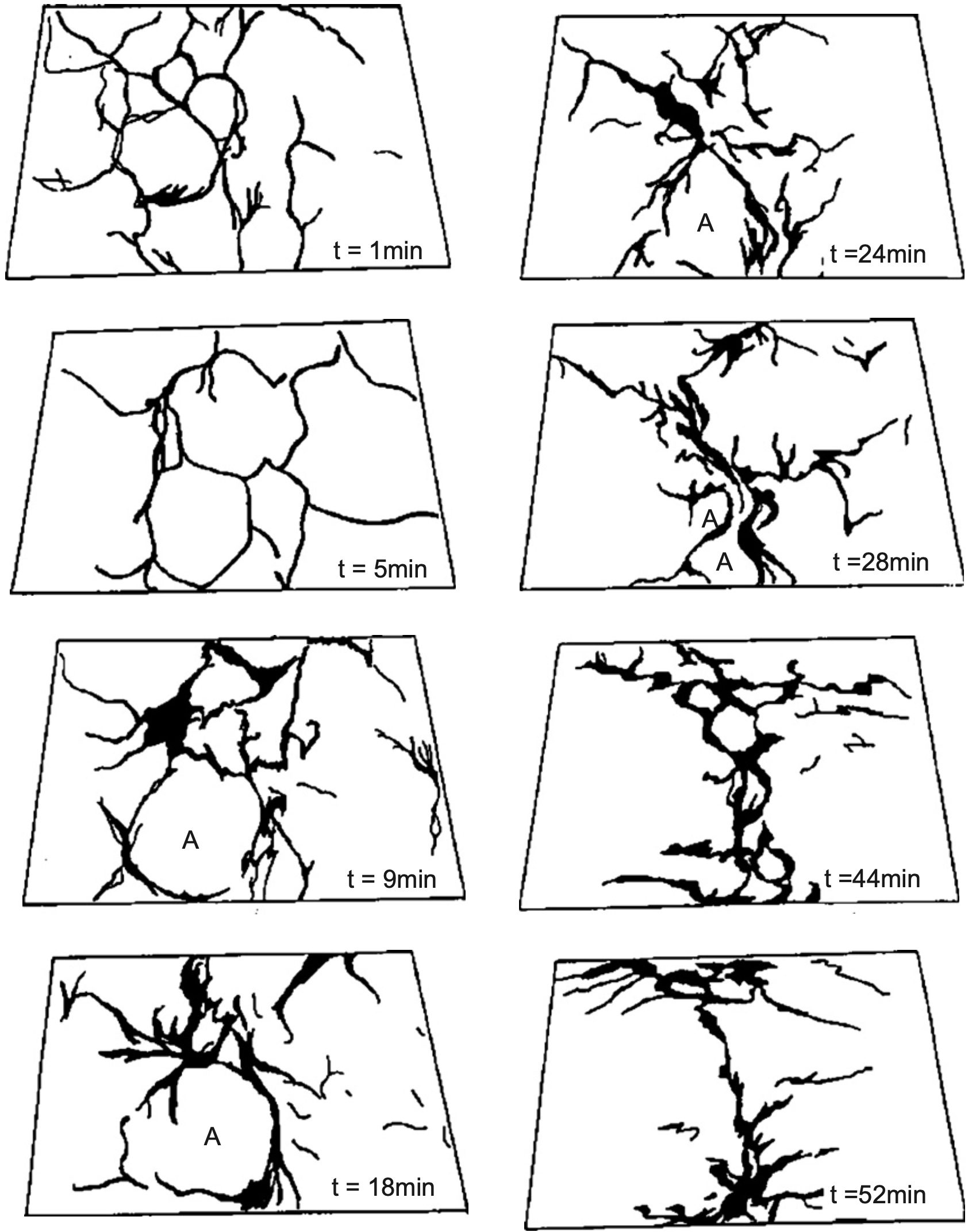}
\label{subfig:iga2006Figure8}
\end{subfigure}
\caption{
(a) Steady convective cells produced in a narrow 2D water tank by a constant injection of bubbles. The whiter the regions are the more concentrated in bubbles. The injection rate of bubbles increases from the first to the third row. (b) Similar observations in numerical simulations: black dots denote the bubble on the first row, while the second row shows the velocity field. Results were obtained for an aspect ratio $L/H=4$ and a Rayleigh number of $2.07\times 10^5$. (c) Sketches of the evolution of a 3D bubble-laden as seen from above, with the time origin corresponding to the start of bubble injection. The dark regions are laden with bubbles. Figure (a) and Figure (c) are respectively adapted from Kimura~(1988) \cite{kimuraCellFormationConvective1988} and Iga and Kimura~(2007) \cite{igaConvectionDrivenCollective2007} \copyright The Japan Society of Fluid Mechanics and IOP Publishing Ltd and reproduced by permission of IOP Publishing. All rights reserved. Figure (b) is adapted from Climent et al.~(1999) \cite{climentLargeScaleSimulationsBubbleInduced1999} with the permission of the American Physical Society.}
\label{fig:bubbleDrivenFlows}
\end{figure}

These previous studies mainly focus on the transient formation of the Rayleigh-Taylor-like instability and essentially analyze the propagation of the front of particles without analyzing the fluid motions that ensue at larger times. If crystals continuously settle in a snow zone, what would be the dynamics of this layer over timescales larger than necessary for crystals to cross through the whole layer? Insight can be gained from experiments of bubble-driven convection: some bubbles are continuously injected at the bottom of a water tank, over the entire cross-section of the tank. Several of these studies focused on the formation of rolls in 2D tanks having a narrow gap between two of their side walls, as observed in experiments of Kimura \cite{kimuraCellFormationConvective1988} (Figure \ref{subfig:KimuraFigure4}) and in numerical simulations by Climent and Magnaudet \cite{climentLargeScaleSimulationsBubbleInduced1999} (Figure~\ref{subfig:ClimentMagnaudetFigure3}). The experiments showed that the rate of injection of bubbles affects the aspect ratio of these rolls, while the size of bubbles controls their ability to couple with fluid motions and cross through the cells when they are sufficiently large \cite{kimuraCellFormationConvective1988}. However, such two-dimensional studies have evidenced very different behaviors compared to observations in 3-D flows. Different regimes exist depending on the bubble injection rate \cite{mezuiBuoyancydrivenBubblyFlows2022}, and while 2D bubble columns tend to show well-structured stacks of rolls, experiments in 3D columns are more chaotic \cite{muddeGravityDrivenBubblyFlows2005}. Experiments of Iga and Kimura \cite{igaConvectionDrivenCollective2007} with a uniform flux of bubbles injected at the bottom of a water tank with a $24\unit{cm} \times 24\unit{cm}$ cross-section have shown that the bubble-driven convection leads to irregular 3D patterns that do not reach a steady state, as shown in Figure \ref{subfig:iga2006Figure8}.
\par
Although bubbles differ from solid crystals by their ability to deform and the possible existence of a slip velocity at their interface, such bubble-driven flows provide valuable information. They support the idea that the motion of crystals in the slurry layer has little chance of being uniform and steady, even when disregarding issues due to bulk crystallization.

These studies confirm the order-of-magnitude impact that collectivity can have on the timescale of crystal segregation in the core and on the estimates of convective velocities that may or may not drive a dynamo. Unfortunately, collective effects are still poorly understood, difficult to investigate experimentally, and challenging to model numerically. Further experiments are required to gain more understanding of these aspects, especially in the context of a global settling-driven convection of solid particles over large timescales, for which the parameter space ($\Pi_\rho$,$\phi$,$Re_p$ and $\mathcal{R}$) remains largely unexplored.

\subsection{Influence of phase change in slurry layers}
\label{subsec:phasechange_slurry}
\subsubsection{\label{subsec:bulkCrystallisationXP}Crystallization}

Slurry layers involve the crystallization of free crystals when rising or falling. In metallurgy, the formation of equiaxed crystals above the front of solidification is associated with the convective motion of the fluid \cite{bw1996}. Experiments of equiaxed crystal solidification (with analog material) have shown the complexity of the multiphase flow associated with crystal growth, as a slurry layer encounters processes of nucleation, fragmentation, sedimentation, and convective fluid motion \cite{bw1995,bw1996,kslw2013}. These experiments \cite{bw1996} and early numerical simulations \cite{bw1995,rdb2003,wl2009} have shown that the microscale dynamics (nucleation, grain growth, sedimentation) cannot be disregarded for an understanding of the macroscale segregation in industrial casting \cite{b2002b}. These dynamics can be modeled only at a small scale or with a few crystals. In the most recent numerical studies, the growth of equiaxed crystals with fluid motions can be modeled using the phase-field method and lattice Boltzmann methods in small computational domains, but with the presence of up to a few hundreds of crystals \cite{stosa2020,yst2021,t2023}. Other experiments with equiaxed crystals have investigated the growth rate of individual falling crystals as a function of the surrounding constitutional supercooling \cite{aacl1999,bcb2007}, or the production rate of crystals through fragmentation of dendrites via convective motion \cite{hll1997,ccr2004,lskw2017}. However, despite the existence of recent techniques to track particles in multiphase flows \cite{anse2019,ante2020}, the fluid motions that naturally emerge in slurry layers make it challenging to perform quantitative measurements, especially in large fluid domains.

In planetary cores, the formation of a slurry layer (the Earth's F-layer or a snow layer in small cores) is thought to be accompanied by a stably stratified layer due to the release of light elements upon crystal growth and the sinking of dense crystal \cite{had2006,wdj2018,wdj2023}. Following the seminal studies done by Loper and Roberts~(1977) \cite{lr1977} and Loper~(1992) \cite{l1992}, theoretical one-dimensional models have been developed \cite{wdj2018,wdj2023} for the Earth's F-layer. They showed that the bulk crystallization of iron crystals and their sedimentation are compatible with the existence of a stratified layer at the bottom of the Earth's core. Numerical models of the thermal evolution of Ganymede or Mars core in a regime of iron snow led to the same conclusion \cite{rbs2018,dp2018}. However, even with several assumptions made (one-dimensional and two-phase flow, Boussinesq approximation), the most advanced model for such a slurry layer requires at least 8 equations and 14 input parameters \cite{wdj2023}.

\begin{figure}
 \centering
 \includegraphics{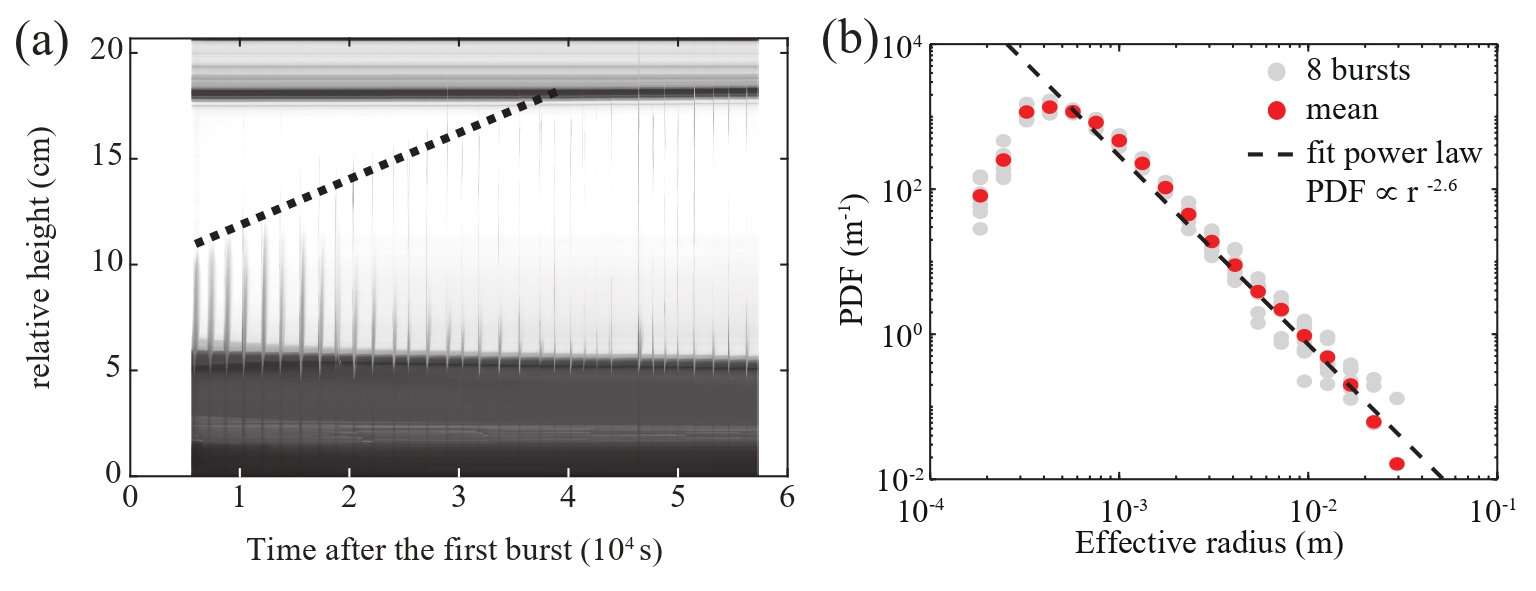}
 \caption{Experimental study of the homogeneous crystallization of a tank of water cooled from below. (a) shows a space-time diagram from a vertical cross-section in the middle of the tank: darker regions correspond to ice. The dynamics exhibit clear alternative behaviors with periods of intense upward snow followed by quiet periods. The black dashed line denotes the maximum height of rising ice crystal i.e. the height of the liquidus. (b) shows the measured probability distribution function of rising snow particle radius, highlighting the large distribution. Panels (a,b) adapted with permission from Huguet et al. (2023) \cite{hld2023}.}
 \label{fig:snow}
\end{figure}

Only a few experiments have been developed to investigate the dynamic of crystallization of a slurry layer in the configuration of iron snow \cite{ante2020,hld2023}. Huguet et al.~(2023) \cite{hld2023} recently investigated the solidification of free ice crystals, their sedimentation, and their melting in a hotter layer (Figure~\ref{fig:snow}). A cyclic pattern was observed of alternating bursts of crystallization and quiet periods with no crystallization (Figure~\ref{fig:snow}a). The bursts of crystallization were due to the formation of large numbers of crystals in a supercooled layer by fragmentation of dendrites. The ice crystals then melted as they rose in hotter regions of the tank. Quiet periods were due to the inhibition of crystallization as the supercooling was reduced in the tank, caused by the release of latent heat of solidification, and the convection induced by the motions of crystals. These experiments revealed the wide range of crystal sizes formed upon solidification in the slurry layer, as shown in Figure~\ref{fig:snow}b. Crystallization in slurry is thought to produce a polydisperse distribution of crystals, which will have consequences on their fall and melting in the iron snow regime. 

\subsubsection{Crystals remelting}

According to the simplistic vision of iron snow previously introduced in section \ref{subsec:scenariosSolidification}, the bottom of the snow zone is delimited by the location where the local temperature is equal to the liquidus temperature; below this position, the temperature is larger than the liquidus and snowflakes remelt. Models of the thermal evolution of small Earth-like planets have time steps of order $10^6\unit{yrs}$ \cite{rbs2015,rbs2018}, which is insufficient to resolve settling and remelting of snowflakes. Therefore, these processes are considered instantaneous, and all snowflakes are assumed to remelt right at the lower edge of the snow zone. In practice, particles have a finite melt rate, and they travel a finite distance before disappearing there exists a `\textit{remelting layer}' between the snow zone and the zone of compositional convection.
\begin{figure}
\centering
\begin{subfigure}[b]{.165\textwidth}
\centering
\caption{}
\includegraphics[width=\textwidth]{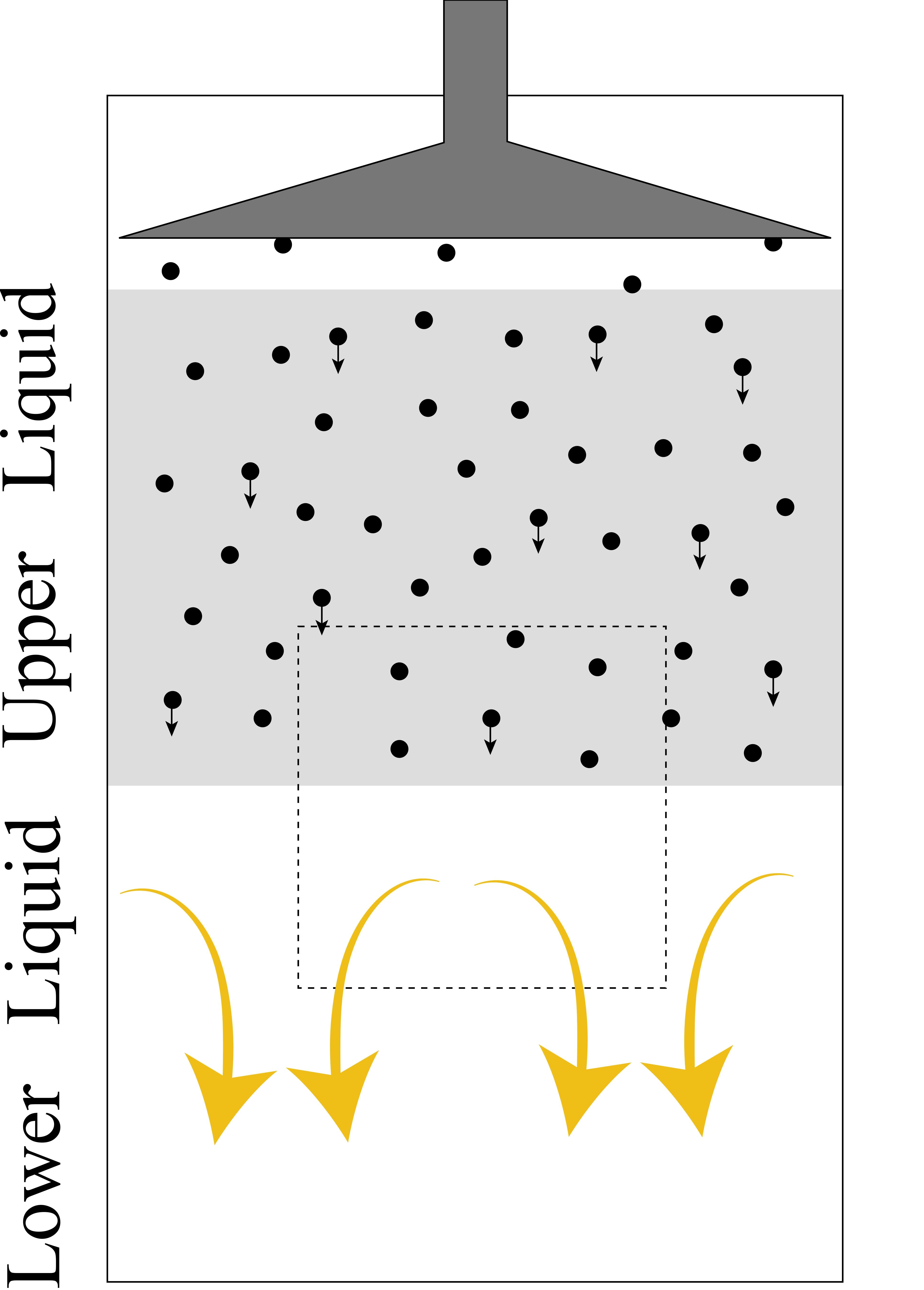}
\label{subfig:sketchOlson}
\end{subfigure}
\begin{subfigure}[b]{.825\textwidth}
\caption{}
\centering
\includegraphics[width=\textwidth]{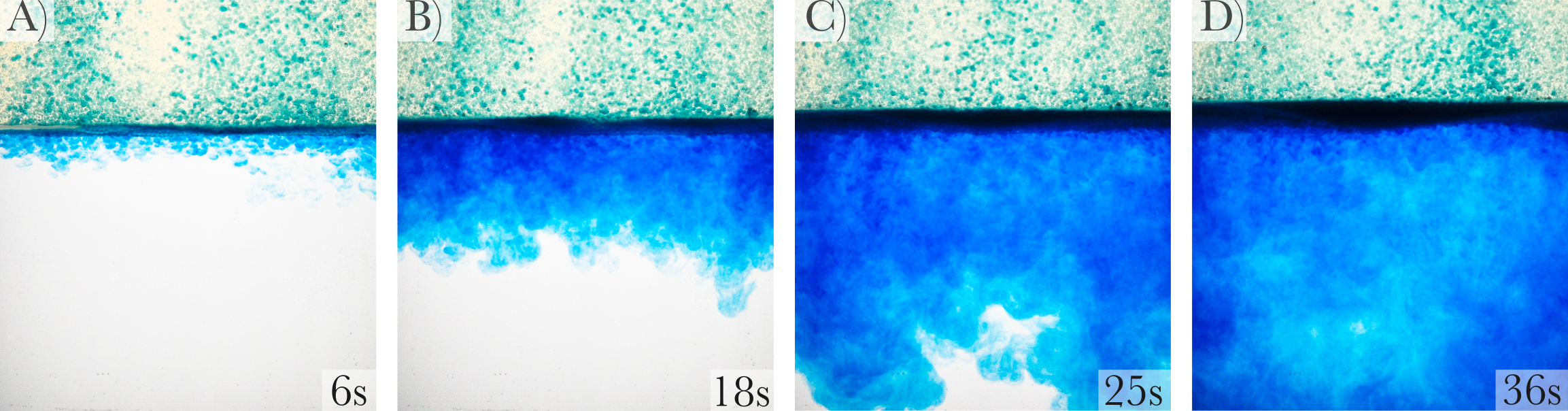}
\label{subfig:convectionOlson}
\end{subfigure}
\caption{(a) Sketch of the setup used by Olson et al.~(2017) \cite{olsonLaboratoryExperimentsRaindriven2017} to model precipitation-driven dynamos. The light grey top layer contains silicone oil and typically stands for the snow zone; the white bottom layer contains fresh water and stands for the convective zone. The dyed aqueous drops that are continuously injected represent settling crystals. (b) Time series of the compositional convection that develops in the lower convective layer. The width of every snapshot is $18\unit{cm}$. We thank Olson et al. for providing panels (a) and (b) which correspond to the experiments used in \cite{olsonLaboratoryExperimentsRaindriven2017}.}
\label{fig:olsonExperiments}
\end{figure}

The remelting layer is usually disregarded. Therefore, models of compositional convection in the deeper core are fully liquid models in which the whole buoyancy is of a fluid nature. As shown in Figure~\ref{subfig:sketchOlson}, Olson et al.~(2017) \cite{olsonLaboratoryExperimentsRaindriven2017} designed an experiment where drops of aqueous solution rained out from a shower head into quiescent air, penetrated through a first layer of silicone oil in which the drops settled -- the analog of the snow zone --, and finally penetrated through a second layer of pure water -- the analog of the convective zone -- where convection developed depending on the buoyancy flux of raindrops, see Figure~\ref{subfig:convectionOlson}. By relating the Rayleigh number which quantifies the convective forcing, to the Reynolds number based on the average velocity of plumes propagating down the second convective layer, the authors showed that this plume velocity verifies an inertial scaling that is independent of viscosity and diffusivity. This observation led Olson et al.~(2017) \cite{olsonLaboratoryExperimentsRaindriven2017} to use diffusion-free dynamo scalings to analyze whether a dynamo could be sustained by precipitation-driven convection on Earth, Mercury, or Ganymede; the details for each case can be found in \cite{olsonLaboratoryExperimentsRaindriven2017}.
\par
Few studies include phase change and particle settling in a large-scale flow. Large-scale studies consider either a fully fluid convection or a settling-driven convection of inert particles. From the physicist and mathematician's viewpoints, melting and dissolving are equivalent \cite{wellsMeltingDissolvingVertical2011,gagliardiThinFilmModeling2018,favierRayleighBenardConvection2019}, so information can be found in the literature on individual particles melting or dissolving as they settle. Most studies have tried to measure and predict the trajectory and shrinking of a single solid sphere falling and dissolving in a homogeneous ambient \cite{riceCompleteDissolutionSpherical1979,riceTranspirationEffectsSolids1982,elperinMassHeatTransfer2005,riceDissolutionSolidSphere2006}. Recently, Huguet et al.~(2020) \cite{hbl2020} studied a settling sphere of $Re_p \sim 10^3-10^4$. They showed that when the particle melts, its trajectory has a lower amplitude of oscillation and a drag coefficient almost twice larger than when it is inert. Aligning with the work of \cite{zmm2019}, Huguet et al.~(2020) \cite{hbl2020} conjecture that is it due to enhanced mixing in the sphere's wake due to the melt's buoyancy. These observations are important for the timescale of crystal segregation in the core; future experiments could clarify the role of the melt liquid by using other materials whose melt liquid is neutrally buoyant, as done in \cite{heSedimentationSingleSoluble2023}. The experiments in \cite{hbl2020} also revealed that when a melting sphere falls in a stratified ambient, it excites internal waves that contain $\sim 1\%$ of the initial potential energy of the sphere for a Brunt-Väisälä frequency $N\sim 0.1-0.2 \unit{Hz}$. How the waves of several crystals might combine in the snow zone, and how they would couple with the crystals' motion remain open questions that require further investigation.
\begin{figure}
\centering
\begin{subfigure}[b]{.28\textwidth}
\caption{$363\unit{\mu m},\ 0.052\unit{g.s^{-1}}$}
\centering
\includegraphics[height=4.8cm]{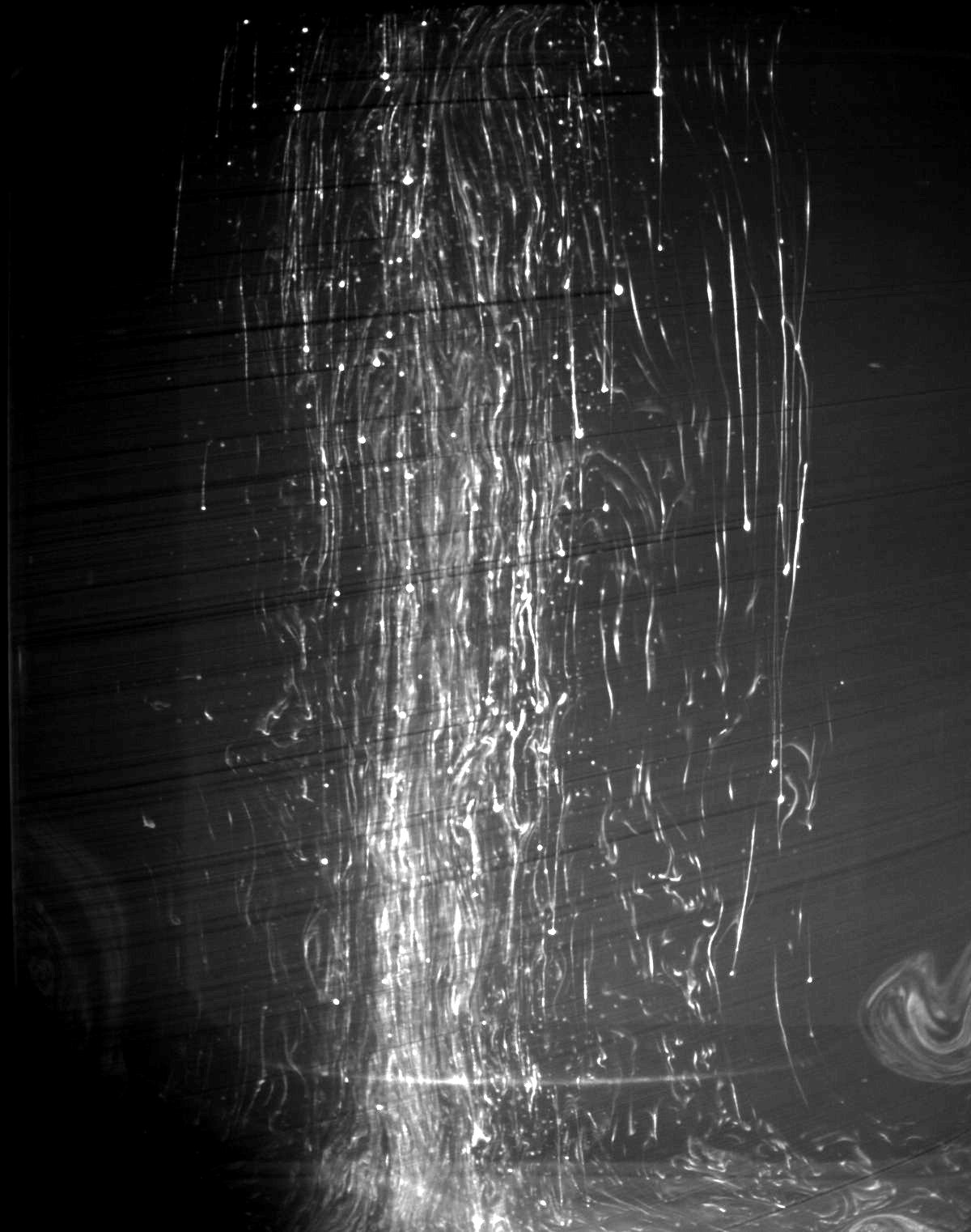}
\label{subfig:snapshots_362microns_52mgpersec}
\end{subfigure}
\begin{subfigure}[b]{.45\textwidth}
\caption{$101\unit{\mu m},\ 0.115\unit{g.s^{-1}}$}
\centering
\includegraphics[height=4.8cm]{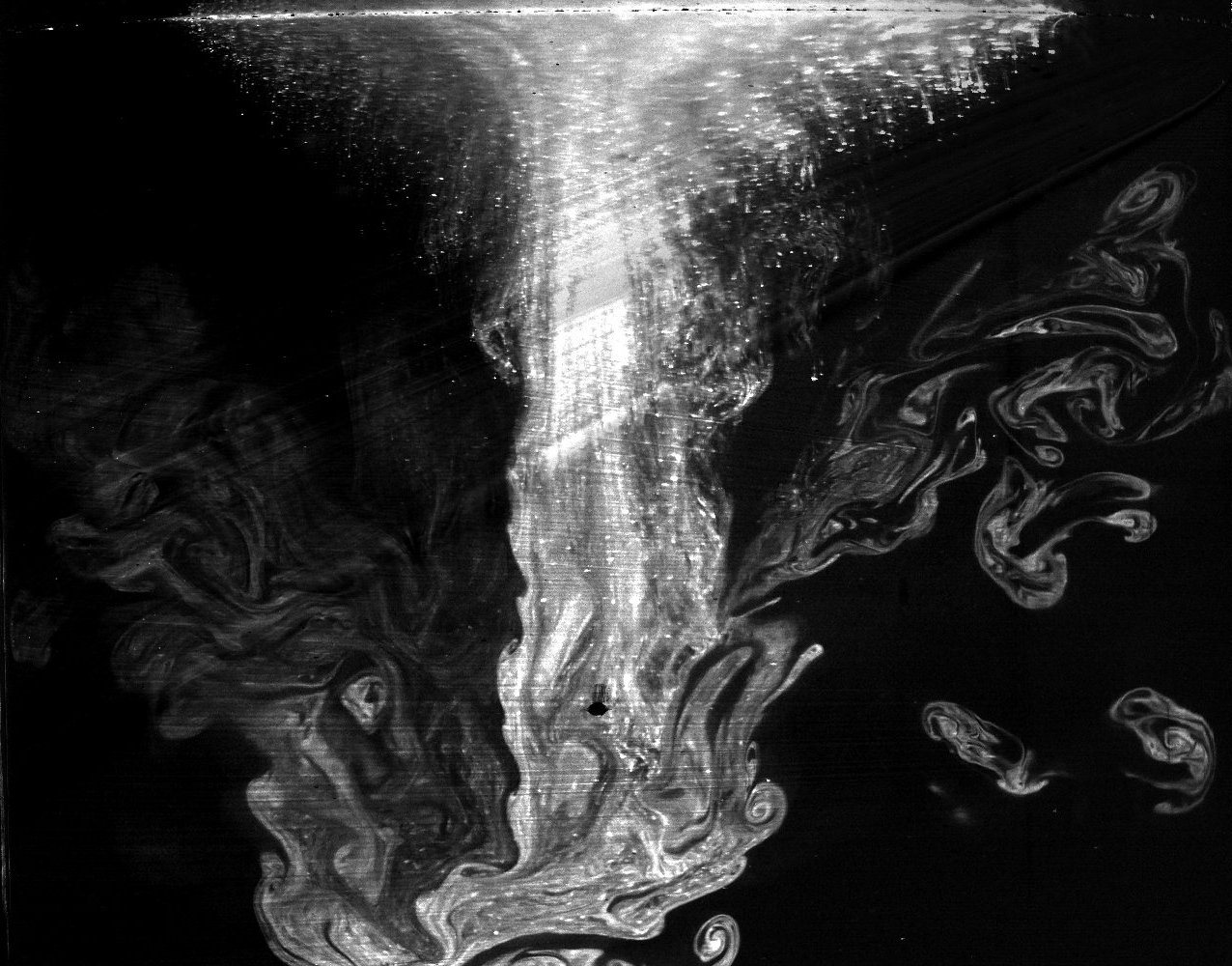}
\label{subfig:snapshots_101microns_115mgpersec}
\end{subfigure}
\begin{subfigure}[b]{.26\textwidth}
\caption{$45\unit{\mu m},\ 0.120 \unit{g.s^{-1}}$}
\centering
\includegraphics[height=4.8cm]{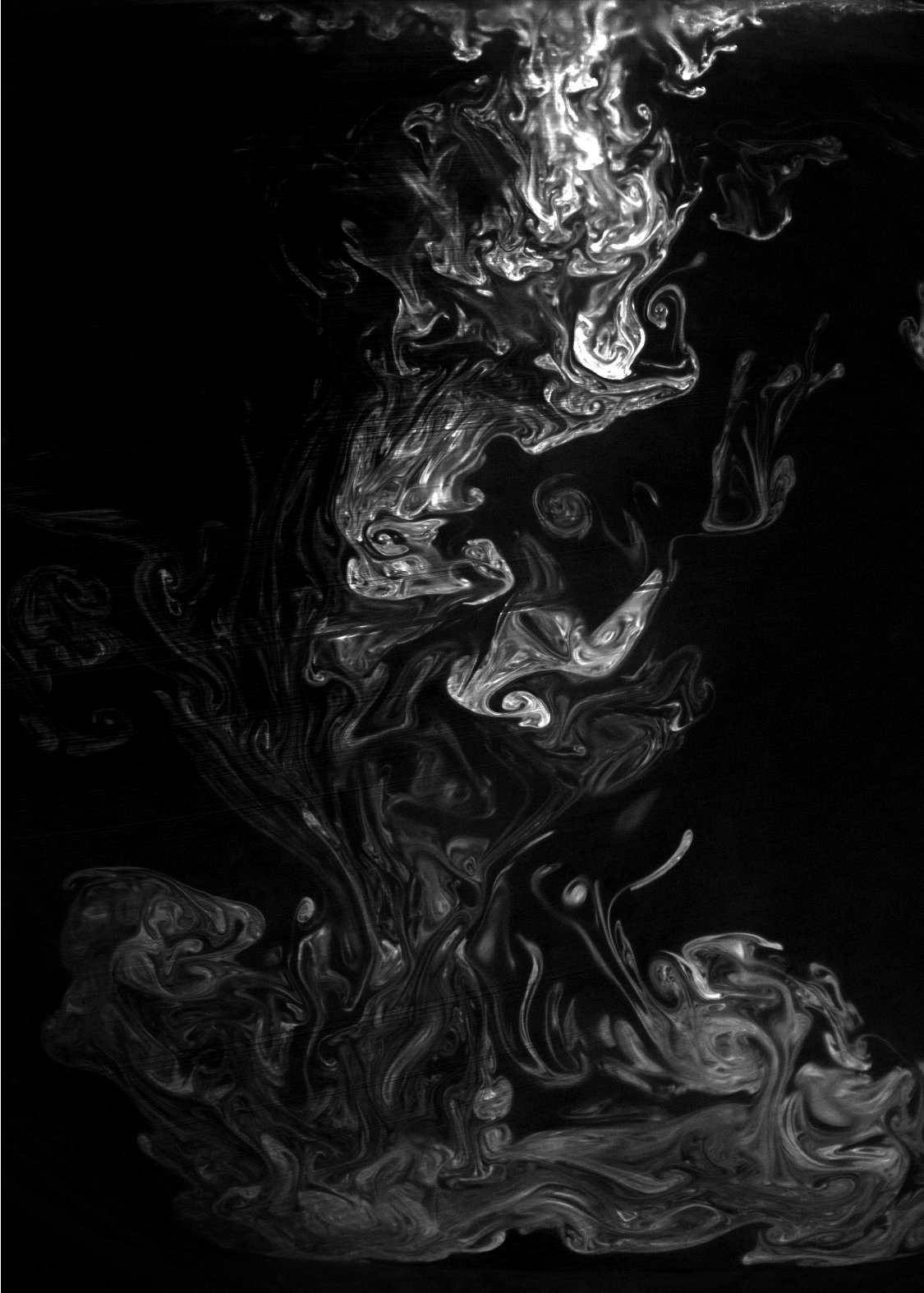}
\label{subfig:snapshots_46microns_120mgpersec}
\end{subfigure}
\par
\caption{Snapshots of sugar plumes of different sugar radii and similar mass rates (see above snapshots). The vertical extent of each snapshot is (a) 38 cm, (b) 23 cm, (c) 38 cm. Snapshots of the experiments presented in Chapter 4 of Kriaa (2024) \cite{kriaaFlowsInducedSettling2023}.}
\label{fig:SugarExperiments}
\end{figure}

Recent experiments were conducted that include the collective aspect of the particles' dynamics while they dissolve (see Chapter 4 in \cite{kriaaFlowsInducedSettling2023}). To do so, homemade dyed sugar grains of controlled size were sieved with a constant mass rate over a cylindrical water tank. A plume of dissolving sugar grains formed over a wide surface area, and a vertical laser sheet enabled to visualize the flow in the midplane. The sugar grains were visible in the laser sheet because they reflected the laser light. In addition, the sugar was cooked like caramel and doped with Rhodamine B, hence the wakes of dissolved sugar were visible in the flow (see Figures~\ref{subfig:snapshots_362microns_52mgpersec}-\ref{subfig:snapshots_46microns_120mgpersec}). By aligning with the usual assumptions of models of core thermal evolution -- a uniform and constant mass flux of monodisperse particles is fed through the bottom of the snow zone (in experiments, through the water-free surface) -- these experiments revealed the importance of the remelting layer and how the size of grains hugely affects its dynamics. First, for an identical mass rate, varying the diameter of sugar grains from $45\unit{\mu m}$ to $363\unit{\mu m}$ controls a flow transition, from a laminar plume that gains intensity over a long transient when forced by the rectilinear fall of large grains (Figure~\ref{subfig:snapshots_362microns_52mgpersec}), to a turbulent lazy plume that emerges faster when forced by fast-dissolving small grains (Figure~\ref{subfig:snapshots_46microns_120mgpersec}). This flow transition is determined by the enhanced forcing imposed by smaller grains, and their ability to nourish a Rayleigh-Taylor-like instability at the plume onset -- both because smaller grains behave more collectively and hardly decouple from fluid motions. Second, the plume velocity explicitly depends on the size of grains through their settling velocity which controls how dilute the grains are in the plume. For these two reasons, the size of grains imprints a lasting signature on the plume, even after full dissolution/melting. Third, the thickness of the remelting layer depends both on the mass flux imposed and on the size of grains (for a mass rate of $0.1 \unit{g/s}$, grains of radius $r_p=45 \unit{\mu m}$ dissolve over a few centimeters whereas grains of radius $r_p=363 \unit{\mu m}$ dissolve deeper than the tank height of $38\unit{cm}$). Since the mass flux of crystals depends on the dynamics of the snow zone that feeds the melting layer with crystals, it reinforces the need to accurately model the fluid motions and the crystal sedimentation in the snow zone.

If crystals are sufficiently large, they may get deep in the core before remelting. As long as they remain large enough for their settling velocity to largely exceed the surrounding flow velocity ($\mathcal{R} \gg 1$), crystals are expected to settle radially inward irrespective of the ambient fluid motions and to inject their buoyancy deep in the core. Only when crystals are sufficiently small can they be entrained in ambient flows? Here again, constraints on the size of crystals are paramount to better model the core evolution and where compositional convection might take place. We mentioned in section \ref{subsec:bulkCrystallisationXP} that polydisperse distributions are expected; it remains an open question to know how polydisperse experiments behave compared to these experiments with monodisperse distributions. Crystals of different sizes settle with different velocities; therefore, they might collide, form clusters, and shield one another from the ambient fluid (see such effects in \cite{k1995}), thus delaying phase change, as already observed in the monodisperse experiments. 

Such idealized experiments show that many open questions remain, even before addressing the geophysically relevant effects of compressibility and magnetism. Such effects will be difficult to tackle experimentally. Yet, to account for the very complex physics involved in phase-change-driven convection, including out-of-equilibrium thermodynamics and crystal size distribution that are inaccessible to Direct Numerical Simulations, laboratory experiments are, at the very least, necessary to test parameterization schemes and to benchmark numerical models that will then address planetary mysteries.

\section{Implications and limitations}
\label{sec:conclusion}
\subsection{Consequences for observations}

A better understanding and modelling of phase-change driven flows in planetary cores is especially relevant in the current context of new observations from recent (e.g., Bepi Colombo) and upcoming (e.g., Juice, Europa Clipper, Psyche) space missions, which will help to unravel the evolution of planetary cores. The dynamics of internal flows will remain inaccessible to observations on many planets, at least within the foreseeable future. The challenge of lab experiments on solidification and melting is therefore to provide large-scale signatures that are closely linked to the microscale physics at play in the core, so as to constrain the core dynamics from planetary-scale observations.

\subsubsection{Dynamo and magnetic observations}

A better understanding of crystal migration could help guide future magnetic observations. The crystals' size affects how they force the flow, decouple from it, and how deep they remelt, thus affecting the characteristic flow velocity $U$ and the size of the convective zone $L$. If they are sufficiently large, the magnetic Reynolds number $Re_m = UL/\eta_m$ can exceed the minimum value required for dynamo action. The more turbulent the flow, the more high-order multipoles can be excited and therefore the more multipolar the magnetic field might be.

However, magnetic observations at the surface of a planet do not only depend on the inner core dynamics. For the same $Re_m$, the magnetic multipoles produced by a deep convective zone (scenario of top-down freezing) decay over a large distance before being measured at the planet's surface, favoring the predominance of the magnetic dipole; conversely, the magnetic field produced by a convective zone near the core-periphery decays over a smaller distance and is prone to show higher-order multipoles \cite{rbs2018}. In addition, solidification can lead to stratification, as expected in Ganymede where top-down freezing leads to the rejection of a sulfur-rich residual liquid rising to the core's top \cite{rbs2015,rbs2018}. Such a stratification filters the magnetic field, reducing higher-order multipoles relative to the magnetic dipole \cite{c2015}. So far, the existence of such a stratification has been deduced from thermodynamical arguments (see e.g., \cite{rbs2018}). Aspects of fluid dynamics remain to be investigated, like the possible erosion of the stratification by the underlying convection, the two-way coupling between the stratification and the flow produced by the settling crystals, and the transport of crystals by inertio-gravity as observed in other astrophysical contexts in astrophysical fluid dynamics \cite{magnanPhysicalPictureAcoustic2024}.

\subsubsection{Seismic observation of the Earth's inner core}

Seismological studies of the Earth's inner core have revealed large and small-scale heterogeneities of the seismic properties (attenuation, anisotropy, or wave velocity), which are thought to be consequences of the evolution of the inner core and then of its crystallization pathway \cite{d2014,wipt2023}. Radial and regional variations of anisotropy are related to the principal orientation of the crystals and their intrinsic properties. Several studies modeling traveling waves in the complex medium have attempted to understand the seismic observations \cite{c2007,ca2013,lcdm2016} and their link to the history of solidification \cite{ca2013}.

As shown in the section \ref{sec:mushylayer}, the structure of the mushy layer originates from the dynamic of the crystallization (growth rate, crystal alignment, chimneys, remelting). Beyond experimental works of crystallization, several experimental studies have attempted to correlate seismic observations to crystallization and dynamics of the Earth's inner core \cite{bghi2000,beo2002,habdll2016}, using ultrasonic waves as analog to seismic waves. Solidification texturing of the mushy layer \cite{b1997} can cause the apparent elastic anisotropy \cite{bghi2000} as crystal elongation aligns with heat flux. A model has been proposed involving convective translation of the inner core \cite{adm2010}: this would produce crystallization in one hemisphere and melting in the other. Huguet et al.~(2016) \cite{habdll2016} have shown that the melting of a mushy layer lowers its attenuation and have argued that the western hemisphere of the inner core melts as it has a weak attenuation. The translation, as such, does not explain the anisotropy nor the innermost inner core. But pre-existing texture \cite{beo2002} or post-solidification deformation \cite{sykh1996,d2012} of the solid mushy layer might alter the solidification texturing, which can perhaps explain anisotropy variation in the inner core or a different structure in the innermost part of the inner core. With the increase in the number and quality of seismic observations, the relation between the mushy layer structure and its seismic properties will deserve more investigation, guided by experimental results to identify the evolution of the Earth's inner core. 

\subsection{Additional questions to address}

On a very long timescale, a growing mushy layer might deform under its weight. Few theoretical and numerical studies with 1D models of compaction \cite{sykh1996,lkc2020} have shown that the length-scale of compaction is relatively small compared to the size of the inner core and that a significant amount of liquid in a compacted mushy layer can be trapped. However, the complexity of extending these models to 2D or 3D renders an experimental approach highly valuable. Compaction in porous medium has been extensively studied through experiments and numerical simulations \cite{ggsve2015,dwzz1994,bblsswz2017}, but not in regard to a mushy layer growing at the inner core boundary. The next set of experiments may investigate the compaction using ``soft'' material like paraffin or lead and applying external compression force or high apparent gravity. We have already seen how convection in a mush increases its solid fraction, without deformation of the solid structure. It would be interesting to see in experiments the second phase of compaction, where the solid fraction increases further because of solid deformation.

Experiments on the crystallization of a core have been performed in many configurations and with different materials (ice, analog salt solution, metals). However, only a few studies have been performed in the configuration of a snow layer \cite{ante2020,hld2023} or of a top-down mushy layer \cite{c2001,ante2020}. Future experiments would have to focus on the stability of a top-down mushy layer and its interaction with the top cooling boundary. In the slurry layer, the processes of crystal growth and fragmentation can be investigated experimentally to predict the size distribution of iron crystals in the snow layer, which is until now unknown.
Indeed, a slurry layer has little chance of being monodisperse: if crystals form anywhere in the bulk, upward or downward settling enables crystals of many sizes to meet and interact in the core; if large pieces of solid iron are delaminated from the core-mantle boundary, they might trigger nucleation in a supercooled ambient, especially if their detachment results in the dispersal of smaller fragments. As we have seen, the crystal size has a drastic influence on convection. Hence, although experiments with monodisperse particle distributions are a necessary step to fully comprehend the physics at play, it is unclear how their results would be affected when moving on to polydisperse distributions. Disentangling all the effects that are coupled might require keeping track of the space-and-time evolution of the size distribution in the flow. 
Two types of experimental hurdles must be overcome. First, surface tension plays a much stronger role in experiments compared to planetary scales. Secondly, deforming solid materials in experiments is difficult, as the available stress is limited by the size of the setup.

The dynamical consequences of planetary rotation remain largely unexplored in experiments. Some studies have considered its influence on miscible plumes \cite{fernandoDevelopmentPointPlume1998,goodmanHydrothermalPlumeDynamics2004,frankAnticyclonicPrecessionPlume2017,sutherlandPlumesRotatingFluid2021}, bubble-laden plumes (e.g., \cite{frankEffectsBackgroundRotation2021}) or particle clouds \cite{ayotteMotionTurbulentThermal1994,helfrichThermalsBackgroundRotation1994,ksfl2022,kriaaInfluencePlanetaryRotation2024}, always in the configuration where the rotation axis and gravity are aligned hence corresponding to flows inside the tangent cylinder. While rotation tends to constrain flow structures to align with and be invariant along the rotation axis, the buoyancy force tends to constrain radial motions. How these constraints compete with one another when gravity and rotation are not aligned, and how the spherical geometry of the core shapes convection, remain two challenging questions for experimentalists. 

The influence of magnetic field on the two-phase flow in mushy or slurry layer has only been sparsely investigated by experiments and theoretical work \cite{bfb1999,bm2020}. Experimental studies usually require metallic phases, which are opaque, making two-phase flow tracking very difficult. Magnetohydrodynamics can be also experimentally studied with a transparent material such as a water-electrolyte solution and an extremely intense magnetic field \cite{apbds2016,apsd2018}. One might think the solidification of such a water-electrolyte solution makes it possible to have direct visualization of two-phase flows affected by a magnetic field.

\subsection{Similar questions in other fields}

Many more communities share common challenges with core dynamicists, starting with geoscientists studying the cooling of magma reservoirs. Crystallizing magmatic flows are prodigiously complex. Crystals can form in the bulk or on the walls \cite{turnerConvectionMixingMagma1986,sparksFluidDynamicsEvolving1997} of, for example, a magma chamber -- a competition that is challenging to model and unravel from geological samples \cite{holnessCrystalSettlingConvection2017,holnessMagmaChambersMush2019}. Since crystal growth injects latent heat, changes the local fluid composition, and modifies its rheology, the fluid properties and buoyancy hugely depend on the distribution and dynamics of crystals \cite{sparksFluidDynamicsEvolving1997,namurCrystallizationInterstitialLiquid2014,holnessMagmaChambersMush2019}. The resulting convection is reflected in the zonation of rocks on the walls and the alignment of crystals with the flow \cite{holnessCrystalSettlingConvection2017,holnessMagmaChambersMush2019}, which depend on the exact convective patterns and chamber geometry \cite{turnerConvectionMixingMagma1986}. Yet, even from the sole point of view of microphysical processes, the interpretation of geological samples is complex: the porous crystals can temporarily retain liquid or move in a saturated environment \cite{sparksFluidDynamicsEvolving1997,namurCrystallizationInterstitialLiquid2014}, which affects their phase change and possibly their density; with polydispersity \cite{kleinInfluenceCrystalSize2017} and aggregation \cite{holnessCrystalSettlingConvection2017}, all these ingredients modify crystal migration, with direct feedback on magma heterogeneity and rheology.
\par
In watercourses, basins, or coasts of cold regions (e.g., \cite{moralesmaquedaPolynyaDynamicsReview2004,martinFrazilIceRivers1981,d1984}), water can be supercooled, leading to the formation of sub-millimetric to millimetric ice crystals called \textit{frazil ice} in turbulent streams. In this context as well, the large number of crystals forming in the turbulent flow cannot be resolved numerically, which is why parameterizations (e.g., \cite{matsumuraLagrangianModellingFrazil2015,rw2018}) or continuum models (e.g., \cite{hermanHighresolutionSimulationsInteractions2020}) are used to model the crystals. Despite numerous laboratory and field studies on frazil ice, the exact nucleation mechanism for crystals remains unknown. Additional questions remain to be solved concerning the influence on frazil growth and aggregation of, notably, initial supercooling, thermal evolution of the water column, turbulence intensity, and double-diffusive mixing in saline waters. In-situ observations (e.g., \cite{d1984}) and laboratory experiments (e.g., \cite{ushioLaboratoryStudySupercooling1993}) have provided valuable information on the morphology of frazil crystals. Yet, despite the careful analyses of ice samples in thin sections or CAT-SCANS \cite{dubeInnerStructureAnchor2014}, a lot remains to be investigated on their aggregation in anchor ice, whether the latter essentially grow in situ or by collecting frazil crystals, what controls the ability of frazil ice to stick to pebbles or vegetation, and how frazil aggregates manage to trap sediments, pollutants or organisms in grease ice (for a review, see \cite{barretteUnderstandingFrazilIce2021}).
\par
A field sharing many challenges with those of the present paper is cloud dynamics, and how it is coupled to cloud microphysics. Although hydrometeors are widely modeled as a continuum in analog experiments (e.g., \cite{shyTurbulentStratifiedInterfaces1991,krugerDynamicsDowndraughtsCold2020,friesInfluenceParticleConcentration2021}) and simulations (e.g., \cite{boffettaEulerianDescriptionDilute2007,realiLayerFormationSedimentary2017,oms2019,hermanHighresolutionSimulationsInteractions2020,lemusModellingSettlingDrivenGravitational2021}), the mathematical grounds of the continuum approach are regularly violated due to excessive particle dilution \cite{chouNumericalStudyParticleinduced2016,melladoTwofluidFormulationCloudtop2010}. Disregarding the particulate nature of the hydrometeors, which imposes a spatially-discretized coupling between the air and the particles \cite{haradaParticlelikeFluidlikeSettling2012,yamamotoNumericalSimulationConcentration2015a}, leads to persistent shortcomings to model how the particle-scale dynamics are coupled to the cloud macrophysics \cite{melladoTwofluidFormulationCloudtop2010,melladoCloudTopEntrainmentStratocumulus2017}: hydrometeors only drag air in their vicinity and they release or absorb latent heat locally. Since drag and rates of phase change depend crucially on the size of hydrometeors \cite{levichPhysicochemicalHydrodynamics1962,fuRadiativeTransfer2006,pruppacherMicrophysicsCloudsPrecipitation2010}, cloud microphysics directly affects latent heating and evaporative cooling in stratocumuli, altering cloud convection and stability, with a direct consequence on cloud coverage and hence cloud feedback \cite{woodStratocumulusClouds2012}. Through precipitation, microphysics also hugely affects cloud aggregation which ``remains one of the largest sources of uncertainty in climate models and thus for reliable projections of climate change'' \cite{benistonGrandChallengesClimate2013}: during extreme precipitation events, the simultaneous fall and evaporation of hydrometeors can cause violent downbursts that spread on the floor as cold winds, favoring the ascent of warm updrafts above them, leading to cloud aggregation near the downbursts \cite{malardelAtmosphericBuoyancydrivenFlows2012,mullerSpontaneousAggregationConvective2022}.
\par
These questions are equally relevant on other planets, for example, concerning helium rains due to H-He demixing in Jupiter (e.g., \cite{ss1977,nettelmannExplorationDoubleDiffusive2015}). Convective storms are common in the gas giants of our solar system \cite{markhamRainyDowndraftsAbyssal2023}. The influence of microphysics on their atmosphere dynamics has been mentioned to explain the heterogeneity of ammonia in Jupiter's atmosphere through the formation of ammonia-rich hailstones \cite{huesoConvectiveStormsAtmospheric2020,guillotStormsDepletionAmmonia2020}, as expected during some long-lasting evaporatively-driven downdrafts in the deep Jovian atmosphere \cite{markhamRainyDowndraftsAbyssal2023}. More work is also required to model clouds on exoplanets. Taking the example of tidally-locked hot Jupiters, the huge temperature contrast between the sub-stellar point and the night side strongly constrains their atmospheric dynamics \cite{parmentier3DMixingHot2013,hellingExoplanetClouds2019}. This creates a planetary-scale 3D flow that affects the dispersal and stability of tracers through vertical mixing, with consequences on cloud coverage if cold traps deplete the atmosphere of some chemicals through precipitation \cite{parmentier3DMixingHot2013}. Here again, the size of the meteors greatly affects the outcome of the scenarios that have been modeled \cite{parmentier3DMixingHot2013}, but the huge amount of chemicals involved further complexify the picture \cite{hellingExoplanetClouds2019}, notably due to the possible change in composition of the particles as they simultaneously grow and migrate in the atmosphere \cite{hellingExoplanetClouds2019,marley2013clouds}.

\bibliographystyle{crunsrt}

\bibliography{biblio}

\end{document}